\documentclass[twocolumn,preprintnumbers,amsmath,amssymb]{revtex4}
\usepackage{txfonts}
\usepackage{amsmath} 
\usepackage{stmaryrd}
\usepackage{amssymb}	
\usepackage{multirow}  
\usepackage{cases}
\usepackage[none]{hyphenat}

\usepackage{graphicx}
\usepackage{dcolumn}
\usepackage{bm}
\usepackage{braket} 
\usepackage{color}
\usepackage{subfigure}

\def\beq{\begin{equation}}
\def\eeq{\end{equation}}
\def\beqa{\begin{eqnarray}}
\def\eeqa{\end{eqnarray}}
\newcommand{\eq}[1]{Eq.(#1)}
\newcommand{\eqs}[1]{Eqs.(#1)}
\newcommand{\fig}[1]{Fig.(#1)}

\newcommand{\abs}[1]{\left| #1 \right|} 
\newcommand{\avg}[1]{\left \langle #1 \right \rangle} 

\newcommand{\ketc}[1]{\left| #1 \right>} 
\newcommand{\matrixel}[3]{\left< #1 \vphantom{#2#3} \right|
 #2 \left| #3 \vphantom{#1#2} \right>} 

\begin{document}
\title{State-dependent atomic excitation by multi-photon pulses propagating along two spatial modes}
\author{Yimin~Wang$^1$, Ji\v{r}\'{i}~Min\'{a}\v{r}$^1$ and Valerio~Scarani$^{1,2}$}
\affiliation{$^1$Centre for Quantum Technologies, National University of Singapore, Singapore\\
$^2$Department of Physics, National University of Singapore, Singapore}

\date{\today}

\begin{abstract}
We investigate the dynamics of a single two-level atom, which interacts with pulses propagating in two spatial-modes (right and left) and frequency-continuum. Using Heisenberg equations of motion, we present the explicit analytical derivations and general formalisms for atomic excitation with two spatial-mode multi-photon pulses in both Fock state and coherent state. Based on those formalisms, we show that perfect atomic excitation by single photon Fock state pulse can only be realized when it is rising-exponentially shaped in the even-mode---a balanced superposition of the two spatial-modes. Single photon from single spatial-mode can only give half of the maximal atomic excitation probability. We also show that the maximum atomic excitation probability with multi-photon pulses in the even-mode is a monotonic function of the average photon number for coherent state, but not for Fock states. Furthermore, we demonstrate that the atomic dynamics can be controlled by the relative phase between the two counter-propagating coherent state pulses incident on the atom, which is not the case with two Fock state pulses.
\end{abstract}
\maketitle
\section{Introduction}
Atom-light interaction is of great interest and fundamental importance in quantum information sciences. In particular, efficient coupling between a single atom and light lies at the heart of scalable quantum networks, where the photon as ``flying qubit'' transfers the information to the ``stationary qubit''---the atom. Recently, the strong coupling between the propagating light and a single natural atom \cite{Tey_2008, Slodicka_2010}, molecule \cite{Wrigge_2008}, quantum dot \cite{Vamivakas_2007} in three-dimensional (3D) geometry or superconducting qubit \cite{Astafiev_2010_sci, Astafiev_2010_prl}, surface plasmon \cite{Chang_2007} in one-dimensional (1D) geometry has been experimentally demonstrated. The 1D systems can reach more efficient coupling compared with the 3D systems, due to the better spatial-mode matching of electromagnetic waves in 1D systems. Theoretically analysis have been done on single and few-photon transport properties in 1D waveguide systems embedding a two-level atom \cite{Shen_2005_prl, Shen_2007_prl} as well as single photon induced entanglement between two atoms in free space \cite{Santos_2012}.

In this paper, we consider a simplified model in which a single two-level atom and propagating pulses interact in the one-dimensional geometry, where we assume perfect spatial-mode matching between the pulse and the atomic emission pattern. It is worth to note that this model can also describe three-dimensional setups, in which the spatial overlap between the angular distribution of the light pulse and atomic dipole pattern is fixed \cite{Wang_2011}. A full quantum-mechanical approach based on time-dependent Heisenberg-Langevin equations is used to provide the general formalism for solving the atomic and two spatial-modes multi-photon dynamics. Following a detailed numerical analysis of atomic dynamics, the dependence of atomic excitation on different spatial-mode photon states are presented and the differences between Fock state wave-packets and coherent state wave-packets are reported in the two spatial-modes 1D configuration.

The paper is organized as follows. In Sec. \ref{sec_model}, we present a theoretical model of the interaction between a two-level atom and two spatial-mode propagating pulses in 1D frequency-continuum. In Sec. \ref{sec_atom}, we derive the general formalism for atomic dynamics with two spatial-mode multi-photon pulses in Fock state and coherent state. In Sec. \ref{sec_sim}, atomic dynamics is studied by numerical simulation on different photon states. The results are briefly summarized in Sec.\ref{sec_con}.

\section{Physical model}
\label{sec_model}
\begin{figure}[b]
\includegraphics[scale=0.35]{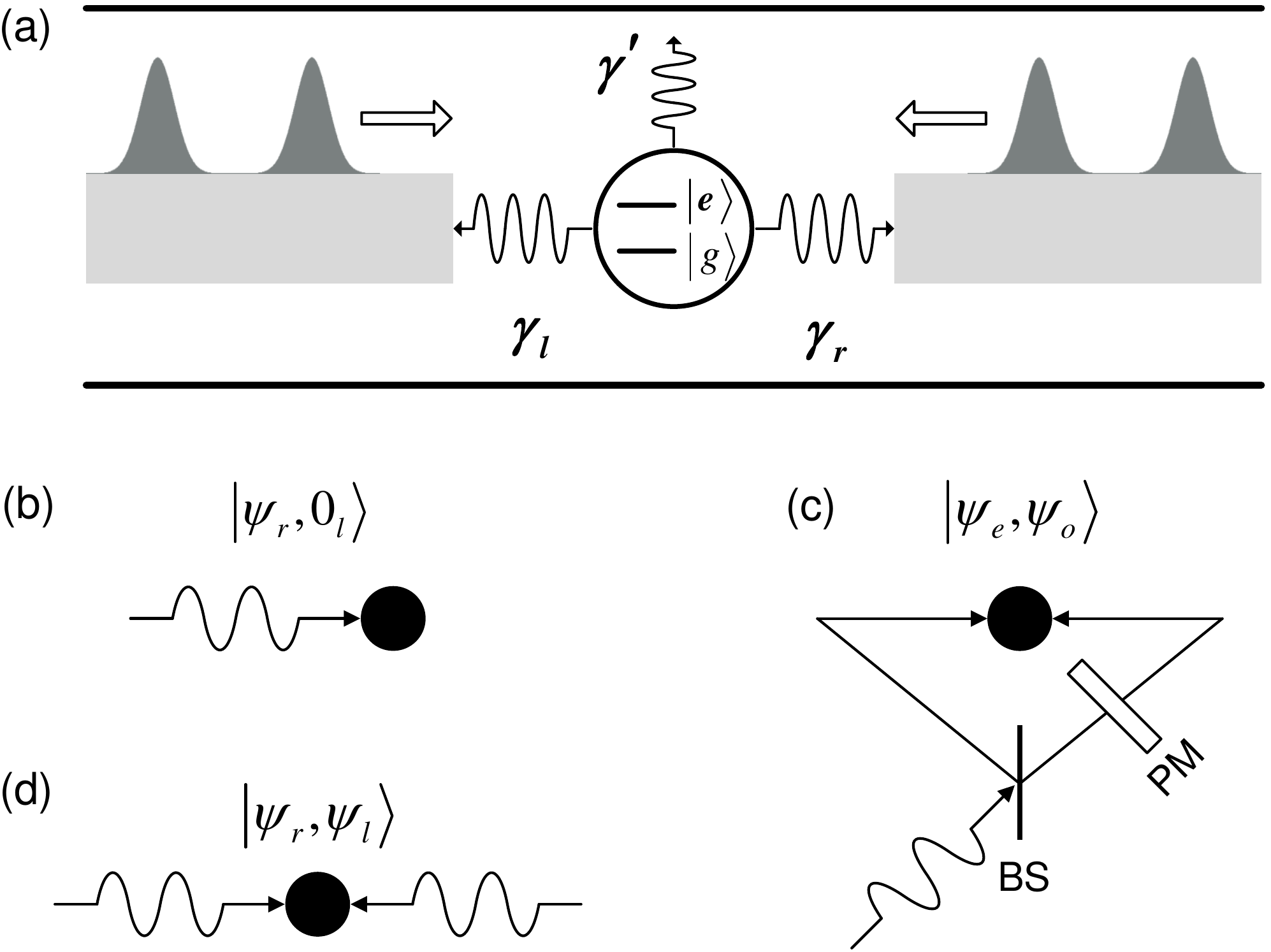}
\caption{(a) Schematic picture of the system: a two-level atom is coupled to the right- and left-propagating pulses with the radiative decay rates $\gamma_r$ and $\gamma_l$, respectively. $\gamma'$ describes the decay rate into the environment. Simplified illustration of the pulse spatial-modes considered in this study: (b) single spatial-mode. (c) even- and odd-parity mode. (d) two distinct spatial-modes. The atom is indicated by the black circle and the arrows indicate the input pulses. BS: beam-splitter. PM: phase modulator.}
\label{fig_2_mode_model}
\end{figure}
As a model system, we consider a two-level atom interacting with one-dimensional photon wave-packets coming from the left, from the right, or from both directions as depicted in \fig{\ref{fig_2_mode_model}} (a). The continuum electric field given in the interaction picture reads \cite{Domokos_2002}
\beqa
\label{eq E+}
\hat{E}^{(+)}(z,t)&=&i \int_0^{\infty} {\rm d} \omega \,A_{\omega,r}\, e^{-i\omega(t - z/c)} \hat{a}_{\omega,r}(t) \nonumber \\
&+&i \int_0^{\infty} {\rm d} \omega \,A_{\omega,l} \, e^{-i\omega(t + z/c)} \hat{a}_{\omega,l}(t)
\eeqa
\noindent where $A_{\omega,j} \propto \sqrt{\omega_j}$ with $ j=r,l $ account for the correct normalization of the electric field in the right- and left-propagating modes. The annihilation operators $\hat{a}_{\omega,j}$ satisfy the usual bosonic commutation relation
\beq
[\hat{a}_{\omega,j}, \hat{a}^{\dag}_{\;\omega',j'}]  = \delta(\omega-\omega') \,\delta_{j,j'}.
\eeq
The atomic dipole operator in the interaction picture reads
\beq
\label{eq d}
\hat{d} = d \left(\hat{\sigma}_- e^{-i\omega_a t} + \hat{\sigma}_+ e^{i\omega_a t} \right),
\eeq
\noindent where $d$ is the value of the dipole momentum, $\omega_a$ is the atomic transition frequency and $\hat{\sigma}_+=\ket{e}\bra{g},  \hat{\sigma}_-=\ket{g}\bra{e},  \hat{\sigma}_z=\ket{e}\bra{e}-\ket{g}\bra{g}$ are the usual two-level atom operators with ground state $\ket{g}$ and excited state $\ket{e}$. We assume that the dipole moment is oriented parallel to the field polarization at the atomic position $z_a$ yielding maximum coupling strength.

The dipole interaction Hamiltonian in the interaction picture and rotating wave approximation is given by \cite{Domokos_2002}
\beqa
\hat{H}_I(t) &=& -i \hbar \sum_{j=r,l} \int d \omega \left[ g_{\omega,j} \,\hat{\sigma}_+ \hat{a}_{\omega,j} \,e^{-i\,(\omega-\omega_a)t}-h.c.\right],
\label{eq_H}
\eeqa
\noindent where the coupling strength is
\beq
\label{eq_g}
g_{\omega,j} =\frac{ d\,A_{\omega,j}}{\hbar} \,e^{\pm i\,\omega \,z_a/c}.
\eeq
The evolution of the system variables is governed by a set of coupled Heisenberg equations
\beqa
\dot{\hat{a}}_{\omega,j} &=& g_{\omega,j}^* \, e^{i(\omega-\omega_a)t}\,\hat{\sigma}_-
\label{eq_da}, \\
\dot{\hat{\sigma}}_{-}&=& -\frac{\gamma'}{2}\hat{\sigma}_{-}+\hat{\zeta}_- +\hat{\sigma}_{z}\sum_{j=r,l}\int d \omega g_{\omega,j} a_{\omega,j} \, e^{-i\,(\omega-\omega_a)t}
\label{eq_dsm}, \\
\dot{\hat{\sigma}}_{z}=&-&\gamma'(\hat{\sigma}_{z}+1)+\hat{\zeta}_z \nonumber \\
&-&2 \sum_{j=r,l}\int d \omega g_{\omega,j} \left[\hat{\sigma}_+ \hat{a}_{\omega,j} \, e^{-i\,(\omega-\omega_a)t}+h.c. \right],
\label{eq_dsz}
\eeqa
\noindent in which the decay term $\gamma'$ and the noise operators $\hat{\zeta}$ are introduced to account for the interaction of the atom with the environment. The explicit form of the noise operator $\hat{\zeta}$ is determined directly in terms of the initial field operators of the environment \cite[p. 273]{Scully_1997}, as discussed in \cite{Wang_2011}.

By integrating Eq.(\ref{eq_da}), both the field operators $\hat{a}_{\omega,j}$ are decomposed into a free field part and a part radiated by the atom \cite [p. 393]{Tannoudji_2004}
\beqa	
\label{eq_da2}
\hat{a}_{\omega,j}(t)=a_{\omega,j}(t_0)+ \int_{t_0}^t d t' g^*_{\omega,j} \hat{\sigma}_-(t')\,e^{i\,(\omega-\omega_a)t'}.
\eeqa
The substitution of \eq{\ref{eq_da2}} back into \eq{\ref{eq_dsm}} and \eq{\ref{eq_dsz}} gives a set of modified optical Bloch equations \cite{Domokos_2002},
\beqa
\dot{\hat{\sigma}}_{-}&=& -\frac{\gamma_0}{2}\hat{\sigma}_{-}+\hat{\zeta}_-+\hat{\sigma}_{z}\sum_{j=r,l}\sqrt{\frac{\gamma_j}{2\pi}}\int d \omega \, e^{-i\,(\omega-\omega_a)t}\,\hat{a}_{\omega,j}(t_0), \nonumber \\
\label{eq_dsm2} \\
\dot{\hat{\sigma}}_{z}&=&-\gamma_0(\hat{\sigma}_{z}+1)+\hat{\zeta}_z \nonumber \\
&-&2 \sum_{j=r,l} \sqrt{\frac{\gamma_j}{2\pi}} \int d \omega \left[\, e^{-i\,(\omega-\omega_a)t}\,\hat{\sigma}_+ \hat{a}_{\omega,j}(t_0) +h.c. \right],\label{eq_dsz2}
\eeqa
\noindent where the free space spontaneous decay rate is made up of three parts: $\gamma_0=\gamma'+\gamma_r+\gamma_l$, the decay into the environment $\gamma'$, and the decay into the right (left) modes $\gamma_r$ ($\gamma_l$), respectively. Using the Weisskopf-Wigner theory \cite[p. 207]{Scully_1997}, the frequency-dependent coupling strengths are approximately constant $g_{\omega,j} \approx g_{\omega_a,j}$, and thus the explicit formula of $\gamma_j$ are given by $\gamma_j = 2 \pi {|g_{\omega_a,j}|}^2$.

In the following, it is convenient to introduce the Fourier-transformed field operators
\beq
\hat{a}_{t,j} = \frac{1}{\sqrt{ 2 \pi}} \int {\rm d} \omega \,\hat{a}_{\omega,j}\,e^{-i\,(\omega-\omega_a)t},
\eeq
\noindent with which the evolution of the atomic operators can be simplified into
\beqa
\dot{\hat{\sigma}}_{-} &=& -\frac{\gamma_0}{2}\hat{\sigma}_{-}+\hat{\zeta}_-+\sum_{j=r,l}\sqrt{\gamma_j}\,\hat{\sigma}_{z}\,\hat{a}_{t,j} \label{eq_dsm3}, \\
\dot{\hat{\sigma}}_{z}&=&-\gamma_0(\hat{\sigma}_{z}+1)+\hat{\zeta}_z -2 \sum_{j=r,l} \sqrt{\gamma_j}  \left[\,\hat{\sigma}_+ \,\hat{a}_{t,j} +h.c. \right].
\label{eq_dsz3}
\eeqa

\section{Atomic excitation with different pulses}
\label{sec_atom}
With the general equations for the atomic operators \eqs{\ref{eq_dsm3}, \ref{eq_dsz3}}, it is now possible to study the dynamics of the atom-pulses system. In this section, we study the probability $P(t)$ of the atom excited by different kinds of photon wave-packets, which is given by the expectation value of the atomic operator
\beq
P(t) = \frac{1}{2}\,\Big( 1 + \bra{\psi(t_0)}\hat{\sigma}_{z}(t)\ket{\psi(t_0)} \Big),
\eeq
\noindent where the initial state of the total system $\ket{\psi(t_0)}=\ket{g,\psi_r,\psi_l,0_s}$ is a product state of the atomic ground state $\ket{g}$, the pulses states $\ket{\psi_r,\psi_l}$ and the environment being in the vacuum state $\ket{0_s}$. As a consequence of the initial vacuum state of the environment, the average values of the noise operators will vanish as $\avg{\hat{\zeta}}=0$.

\subsection{Multi-photon Fock state pulse}
The j-th spatial-mode photon wave-packet creation operator is defined as \cite[p. 243]{Loudon_2000}
\beq
\hat{A}^\dag_j= \int {\rm d} t \, \xi_j(t) \,\hat{a}^\dag_{t,j} = \int {\rm d} \omega \,f_j(\omega) \,\hat{a}^\dag_{\omega,j},
\eeq
\noindent where $\xi_j(t)$ is the temporal shape of the wave-packet and $f_j(\omega)$ is the spectral distribution function. They are related by the Fourier transform
\beq
\xi_j(t) =\frac{1}{\sqrt{ 2 \pi}} \int {\rm d} \omega \,f_j(\omega)\,e^{-i(\omega-\omega_0) t},
\eeq
\noindent with $\omega_0$ being the carrier frequency of the wave-packet. Since we are considering the on resonance interaction of the atom and the pulse, we put $\omega_0=\omega_a$.
The amplitudes are normalized according to
\beq
\int {\rm d} t \,\abs{\xi_j(t)}^2 = \int {\rm d} \omega \,\abs{f_j(\omega)}^2 =1.
\eeq
The j-th spatial-mode photon Fock state is expressed in terms of the wave-packet operator
\beq
\ket{n_j}=\frac{1}{\sqrt{n_j!}}{\Big(\hat{A}^\dag_j \Big)}^{n_j}\, \ket{0},
\eeq
and the corresponding photon number operator is
\beq
\hat{n}_j = \int {\rm d} t \, \hat{a}^\dag_{t,j} \,\hat{a}_{t,j} = \int {\rm d} \omega \,\hat{a}^\dag_{\omega,j} \,\hat{a}_{\omega,j}.
\eeq
In this way, the two spatial-modes photon Fock state is given by
\beqa
\ket{n_r,n_l} &=& \frac{1}{\sqrt{n_r!\,n_l!}}{\Big(\hat{A}^\dag_r \Big)}^{n_r}\,{\Big(\hat{A}^\dag_l \Big)}^{n_l}\, \ket{0_r, 0_l},
\label{eq_2modef}
\eeqa
\noindent with the total photon number operator being
\beq
\hat{n}=\hat{n}_r+\hat{n}_l.
\eeq

\subsection{Atomic excitation with Fock state pulses}
With the help of \eqs{\ref{eq_dsm3}, \ref{eq_dsz3}}, the expectation value of atomic operator $\avg{\hat{\sigma}_{z}(t)}$ with the initial state of the total system $\ket{g,n_r,n_l,0_s}$ will further depend on different state-dependent values of $\avg{\hat{\sigma}_{\pm}(t)}$. Let us define the following variables
\begin{subequations}
\begin{align}
X_{n_r n_l,\,n_r n_l}&= \matrixel{g,n_r,n_l,0_s}{\hat{\sigma}_z(t)}{g,n_r,n_l,0_s}, \\
Y_{n_r n_l,\,n_r n_l}&= \matrixel{g,n_r,n_l,0_s}{\hat{\sigma}_-(t)}{g,n_r,n_l,0_s}, \\
Z_{n_r n_l,\,n_r n_l}&= \matrixel{g,n_r,n_l,0_s}{\hat{\sigma}_+(t)}{g,n_r,n_l,0_s}.
\end{align}
\label{eq_dsfnxyz}
\end{subequations}
\noindent Using the property of the action of the field operator on the Fock state
\beq
\label{eq_af}
\hat{a}_{t,j}\,\ket{n_j} = \sqrt{n_j}\,\xi_j(t) \,\ket{{n_j-1}},
\eeq
and $\avg{\hat{\zeta}}=0$ as the environment is initially in the vacuum state, one finds a set of recursive differential equations,

\begin{subequations}
\begin{align}
\dot{X}_{n_r n_l,\,n_r n_l}=&-\gamma_0 \left(X_{n_r n_l,\,n_r n_l}+1\right) \nonumber \\
&-2 \sqrt{\gamma_r\,n_r} \,\xi_r(t) \,\left(Y_{(n_r-1) n_l,\,n_r n_l}+Z_{n_r n_l,\,(n_r-1) n_l}\right) \nonumber \\
&-2 \sqrt{\gamma_l\,n_l} \,\xi_l(t)  \,\left(Y_{n_r (n_l-1),\,n_r n_l}+Z_{n_r n_l,\,n_r (n_l-1)}\right), \nonumber \\
\label{eq_dsxfn1} \\
\dot{Y}_{(n_r-1) n_l,\,n_r n_l}=&-\frac{\gamma_0}{2} Y_{(n_r-1) n_l,\,n_r n_l} \nonumber \\
&+\sqrt{\gamma_r\,n_r} \,\xi_r(t)\, X_{(n_r-1) n_l,\,(n_r-1) n_l} \nonumber \\
&+\sqrt{\gamma_l\,n_l} \,\xi_l(t)\, X_{(n_r-1) n_l,\,n_r (n_l-1)},
\label{eq_dsyfn1} \\
\dot{Y}_{n_r (n_l-1),\,n_r n_l}=&-\frac{\gamma_0}{2}Y_{n_r (n_l-1),\,n_r n_l} \nonumber \\
&+\sqrt{\gamma_r\,n_r} \,\xi_r(t)\, X_{n_r (n_l-1),\,(n_r-1) n_l} \nonumber \\
&+\sqrt{\gamma_l\,n_l} \,\xi_l(t)\, X_{n_r (n_l-1),\,n_r (n_l-1)},
\label{eq_dsyfn2} \\
\dot{Z}_{n_r n_l,\,(n_r-1) n_l}=&-\frac{\gamma_0}{2}Z_{n_r n_l,\,(n_r-1) n_l} \nonumber \\
&+\sqrt{\gamma_r\,n_r} \,\xi_r(t)\, X_{(n_r-1) n_l,\,(n_r-1) n_l} \nonumber \\
&+\sqrt{\gamma_l\,n_l} \,\xi_l(t)\, X_{n_r (n_l-1),\,(n_r-1) n_l},
\label{eq_dszfn1} \\
\dot{Z}_{n_r n_l,\,n_r (n_l-1)}=&-\frac{\gamma_0}{2}Z_{n_r n_l,\,n_r (n_l-1)}  \nonumber \\
&+\sqrt{\gamma_r\,n_r} \,\xi_r(t)\, X_{(n_r-1) n_l,\,n_r (n_l-1)} \nonumber \\
&+\sqrt{\gamma_l\,n_l} \,\xi_l(t)\, X_{n_r (n_l-1),\,n_r (n_l-1)},
\label{eq_dszfn2} \\
& \vdots \nonumber \\
X_{0_r 0_l,\,0_r 0_l}=&-1,
\label{eq_dsxfn2}
\end{align}
\label{eq_dsfn}
\end{subequations}
\noindent with the initial conditions
\begin{subequations}
\begin{align}
X_{n_r n_l,\,n_r n_l}(t_0)&=-1, \\
Y_{n_r n_l,\,n_r n_l}(t_0)&=Z_{n_r n_l,\,n_r n_l}(t_0)= 0.
\end{align}
\label{eq_dsfn0}
\end{subequations}

\subsection{Multi-photon coherent state pulse}
The j-th spatial-mode coherent state is defined as
\beq
\ket{\alpha_j}=\hat{D}_j(\alpha_j)\, \ket{0},
\eeq
\noindent with the displacement operator
\beq
\hat{D}_j(\alpha_j)=\exp \Big[\alpha_j \hat{A}^\dag_j- \alpha_j^* \hat{A}_{j} \Big].
\eeq
\noindent The mean photon number $\bar{n}_j$ in the j-th spatial-mode is given by
\beq
\bar{n}_j = \matrixel{\alpha_j}{\hat{A}^\dag_j\hat{A}_j}{\alpha_j} = \abs{\alpha_j}^2.
\eeq
The two spatial-modes coherent state reads
\beqa
\ket{\alpha_r,\alpha_l}&=&\hat{D}_r(\alpha_r)\,\hat{D}_l(\alpha_l) \,\ket{0_r, 0_l} \nonumber \\
&=&\prod_{j=r,l} \exp \Big[\alpha_j \hat{A}^\dag_j - \alpha_j^* \hat{A}_j \Big] \,\ket{0_r, 0_l},
\label{eq_2modec}
\eeqa
\noindent and the total mean photon number $\bar{n}$ is
\beq
\bar{n} = \bar{n}_r+\bar{n}_l = \abs{\alpha_r}^2+\abs{\alpha_l}^2.
\eeq
\subsection{Atomic excitation with coherent state pulses}
Similarly, we can study the atomic evolution with two counter-propagating pulses in coherent state $\ket{\alpha_r, \alpha_l}$, which have the property that
\begin{subequations}
\begin{align}
\hat{a}_{t,r}\,\ket{\alpha_r} &= \alpha_r\,\xi_r(t) \,\ket{\alpha_r},  \\
\hat{a}_{t,l}\,\ket{\alpha_l} &= \alpha_l\,\xi_l(t) \,\ket{\alpha_l}.
\end{align}
\label{eq_ac}
\end{subequations}
\noindent Since we are interested only in the interference effect between two spatial-mode fields, the above expressions can be written with the following replacements,
\beq
\alpha_r =\sqrt{\bar{n}_r} , \;\;\;\:
\alpha_l =\sqrt{\bar{n}_l}\cdot e^{i \phi},
\eeq
\noindent where $\phi$ is the initial relative phase between the $\ket{\alpha_r}$ and $\ket{\alpha_l}$.

Again, by taking the average values of \eqs{\ref{eq_dsm3}, \ref{eq_dsz3}} on the initial state $\ket{\psi(t_0)}=\ket{g,\alpha_r, \alpha_l,0_s}$, we have the following differential equations
\begin{subequations}
\begin{align}
\dot{X}_{\alpha_r \alpha_l,\,\alpha_r \alpha_l}=&-\gamma_0 \left(X_{\alpha_r \alpha_l,\,\alpha_r \alpha_l}+1\right) \nonumber \\
&-2 \sqrt{\gamma_r\,\bar{n}_r} \,\xi_r(t) \,\left(Y_{\alpha_r \alpha_l,\,\alpha_r \alpha_l}+Z_{\alpha_r \alpha_l,\,\alpha_r \alpha_l}\right)\nonumber \\
&-2 \sqrt{\gamma_l\,\bar{n}_l} \,\xi_l(t)\,\left(e^{-i \phi}\,Y_{\alpha_r \alpha_l,\,\alpha_r \alpha_l}+e^{i \phi} \,Z_{\alpha_r \alpha_l,\,\alpha_r \alpha_l}\right),
\label{eq_dsxcn} \\
\dot{Y}_{\alpha_r \alpha_l,\,\alpha_r \alpha_l}=&-\frac{\gamma_0}{2}Y_{\alpha_r \alpha_l,\,\alpha_r \alpha_l} \nonumber \\
&+\left(\sqrt{\gamma_r\,\bar{n}_r} \,\xi_r(t)+e^{i \phi} \sqrt{\gamma_l\,\bar{n}_l} \,\xi_l(t)\right)\, X_{\alpha_r \alpha_l,\,\alpha_r \alpha_l},
\label{eq_dsycn} \\
\dot{Z}_{\alpha_r \alpha_l,\,\alpha_r \alpha_l}=&-\frac{\gamma_0}{2}Z_{\alpha_r \alpha_l,\,\alpha_r \alpha_l} \nonumber \\
&+\left(\sqrt{\gamma_r\,\bar{n}_r} \,\xi_r(t) + e^{-i \phi} \sqrt{\gamma_l\,\bar{n}_l} \,\xi_l(t) \right)\, X_{\alpha_r \alpha_l,\,\alpha_r \alpha_l},
\label{eq_dszcn}
\end{align}
\label{eq_dscn}
\end{subequations}
\noindent with
\begin{subequations}
\begin{align}
X_{\alpha_r \alpha_l,\,\alpha_r \alpha_l}&= \matrixel{g,\alpha_r,\alpha_l,0_s}{\hat{\sigma}_z(t)}{g,\alpha_r,\alpha_l,0_s},  \\
Y_{\alpha_r \alpha_l,\,\alpha_r \alpha_l}&= \matrixel{g,\alpha_r,\alpha_l,0_s}{\hat{\sigma}_-(t)}{g,\alpha_r,\alpha_l,0_s},  \\
Z_{\alpha_r \alpha_l,\,\alpha_r \alpha_l}&= \matrixel{g,\alpha_r,\alpha_l,0_s}{\hat{\sigma}_+(t)}{g,\alpha_r,\alpha_l,0_s},
\end{align}
\label{eq_dscnxyz}
\end{subequations}
and the initial conditions
\begin{subequations}
\begin{align}
X_{\alpha_r \alpha_l,\,\alpha_r \alpha_l}(t_0)&=-1, \\
Y_{\alpha_r \alpha_l,\,\alpha_r \alpha_l}(t_0)&=Z_{\alpha_r \alpha_l,\,\alpha_r \alpha_l}(t_0)= 0.
\end{align}
\label{eq_dscn0}
\end{subequations}
\subsection{Even- and odd-parity modes}
For better understanding of the atomic dynamics with two spatial-mode pulses, let us introduce even- and odd-parity mode operators as the combination of the right- and left-propagating modes
\beq
\label{eq_aeo}
\hat{b}^{\dag}_{\omega,e}=\frac{\hat{a}^\dag_{\omega,l}+\hat{a}^\dag_{\omega,r}}{\sqrt{2}} , \;\;\;\:
\hat{b}^{\dag}_{\omega,o}=\frac{\hat{a}^\dag_{\omega,l}-\hat{a}^\dag_{\omega,r}}{\sqrt{2}} .
\eeq
The Fock state pulse in the even-mode is thus given by
\beq
\ket{n_e}=\frac{1}{\sqrt{n_e!}}{\Big(\hat{B}^\dag_e \Big)}^{n_e}\, \ket{0_r, 0_l} ,
\label{eq_even_f}
\eeq
where the wave-packet creation operator for the even-mode is
\beq
\label{eq_even}
\hat{B}^\dag_e= \int {\rm d} t \, \xi_e(t) \,\hat{b}^\dag_{t,e} = \int {\rm d} \omega \,f_e(\omega) \,\hat{b}^\dag_{\omega,e}.
\eeq
Accordingly, the coherent state pulse in the even-mode is
\beqa
\ket{\alpha_e}&=&\hat{D}_e(\alpha_e)\, \ket{0_r, 0_l} \nonumber \\
&=&\exp \Big[\alpha_e \hat{B}^\dag_e- \alpha_e^* \hat{B}_{e} \Big] \, \ket{0_r, 0_l} \nonumber \\
&=&\exp \Big[\frac{\alpha_e \hat{A}^\dag_r- \alpha_e^* \hat{A}_{r}}{\sqrt{2}} \Big] \, \exp \Big[\frac{\alpha_e \hat{A}^\dag_l- \alpha_e^* \hat{A}_{l}}{\sqrt{2}}\Big] \, \ket{0_r, 0_l} \nonumber \\
&=&\ketc{\frac{1}{\sqrt{2}}\alpha_e,\frac{1}{\sqrt{2}}\alpha_e}.
\label{eq_even_co}
\eeqa
For a specific case of single-photon in the even-mode, we have
\beq
\ket{1_e} = \hat{B}^\dag_e \, \ket{0_r, 0_l} = \frac{1}{\sqrt{2}}\,\Big(\ket{1_r, 0_l}+\ket{0_r, 1_l}\Big),
\eeq
for Fock state pulse and
\beq
\ket{\alpha_e}=\ketc{\frac{1}{\sqrt{2}} \alpha_e ,\frac{1}{\sqrt{2}} \alpha_e}, \;\;\;\: \bar{n}_e=\abs{\alpha_e}^2=1
\eeq
for coherent state pulse.

The interaction Hamiltonian \eq{\ref{eq_H}} can be rewritten as
\beqa
\label{eq_Heo}
\hat{H}_I(t) &=& -i \hbar \sqrt{\frac{\gamma_r}{\pi}}\, \int d \omega \left[ \hat{\sigma}_+ \hat{b}_{\omega,e} \,e^{-i\,(\omega-\omega_a)t}-h.c.\right],
\eeqa
\noindent where we have used the property of $g_{\omega,j} \approx g_{\omega_a,j}=\sqrt{\frac{\gamma_r}{2\pi}}=\sqrt{\frac{\gamma_l}{2\pi}}$ under Weisskopf-Wigner approximation and the assumption of equivalent decay into the right and left channels. It is clear from \eq{\ref{eq_Heo}} that only photons in the even-mode interact with the atom and thus contribute to the atomic inversion process, and the odd-mode photons are interaction-free.

\section{Simulation for different photon states}
\label{sec_sim}
In this section, we will use the above formalism for Fock state pulses \eq{\ref{eq_dsfn}} and coherent state pulses \eq{\ref{eq_dscn}} to study the interaction between the two spatial-mode pulses incident on the atom. For this purpose, we consider three distinct cases:

(i) {\it single-photon excitation}---total photon excitation number is one for different spatial-modes;

(ii) {\it multi-photon excitation}---arbitrary $n$-photon in the even-mode;

(iii) {\it two spatial-mode pulses interference}--- $n$-photon in two distinct spatial-modes.

Numerical simulation is done with a specific pulse temporal shape---rectangular shape
\beq
\xi(t)=\Bigg\{
\begin{array}{cc}
 \sqrt{\frac{\Omega}{2}},  & \text{for} \,-\frac{2}{\Omega} \leq t \leq 0 \\
 0,  & else
\end{array},
\eeq
\noindent with $\Omega$ being the bandwidth of the pulse. However, in order to show the coherence of our formalisms with the time-reversed single-photon spontaneous emission process (see below) we will use the rising exponential temporal shape
\beq
\xi(t)=\Bigg\{
\begin{array}{cc}
 \sqrt{\Omega} \exp\left(\frac{\Omega}{2}\,t\right), & \text{for} \,t < 0 \\
 0,  &  \text{for} \, t > 0
\end{array},
\eeq
\noindent for the single-photon excitation case.
For simplicity, we assume a symmetric two spatial-mode structure---no decay to the environment $\gamma'=0$ and equivalent decay to the right and left channel $\gamma_r=\gamma_l=\gamma_0/2$.

\subsection{Single-photon excitation}
In this part, we consider atomic excitation with total photon number one for both Fock state and coherent state pulses in two cases: (a) single-photon in single spatial-mode---$\ket{1_r, 0_l}$ and $\ket{\alpha_r, 0_l}$ with $\bar{n}_r=1$, as seen in \fig{\ref{fig_2_mode_model}}(b). (b) single-photon in the even-mode, $\ket{1_e}$ and $\ket{\alpha_e}$ with $\bar{n}_e=1$, as in \fig{\ref{fig_2_mode_model}}(c).

\begin{figure}[h!]
\includegraphics[scale=0.3]{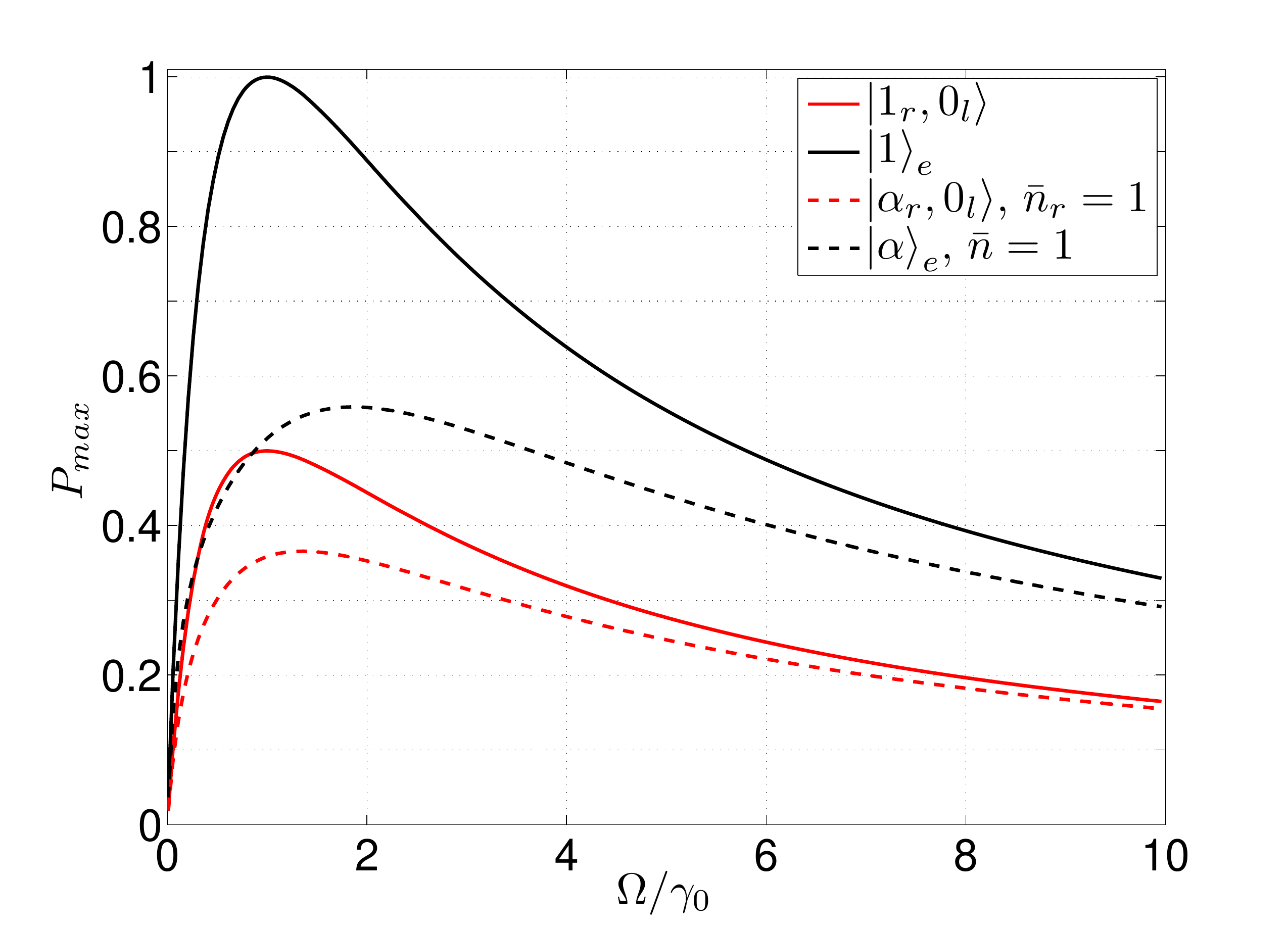}
\caption{(Color online) Dependence of maximum excitation probability $P_{max}$ on pulse bandwidth with initial rising exponential shape for single-photon Fock state (solid curves) and coherent state (dashed curves) in the right-propagating mode (red curves) and the even-mode (black curves), respectively. Full atomic excitation by single-photon pulse can only be realized when it is rising-exponential shaped in the even-parity mode Fock state $\ket{1}_e$. The excitation probability is bounded by $0.5$ if the single-photon only occupies single spatial-mode $\ket{1_r,0_l}$.}
\label{fig_pmax_bwi_exp}
\end{figure}

In \fig{\ref{fig_pmax_bwi_exp}}, we show the dependence of the maximum achievable excitation probability $P_{max}$ on the pulse bandwidth $\Omega$ for different photon states with a rising exponential temporal shape. We find out that for Fock state pulse $\ket{1_r, 0_l}$ (red solid curve), the maximum excitation probability $P_{max}=0.5$, and it goes up to $P_{max}=1$ for a single-photon in the even-parity mode $\ket{1_e}$ (black solid curve), with the same optimum pulse bandwidth $\Omega=\gamma_0$. This is not surprising since the photon in the single spatial-mode (left or right) has half probability of being in the odd-mode, which doesn't interact with the atom. Another explanation is that the photon in single spatial-mode cannot cover the whole dynamics in this two spatial-mode description of the atom-pulses interaction, because of the atomic relaxation into the other channel. This observation simply agrees with the time-reversed spontaneous emission argument. The sufficient condition for a two-level atom to be fully excited by a single-photon is that the single-photon has to be rising-exponentially shaped in Fock state in addition to perfect spatial-mode matching with the atomic dipole \cite{Wang_2011}. Full excitation of a single two-level atom by a single photon with arbitrary temporal shape is possible when the atomic decay rate can be modified by shaping a fast moving mirror, as proposed in \cite{Wang_2012_PRA}.

\begin{figure}[h!]
\includegraphics[scale=0.3]{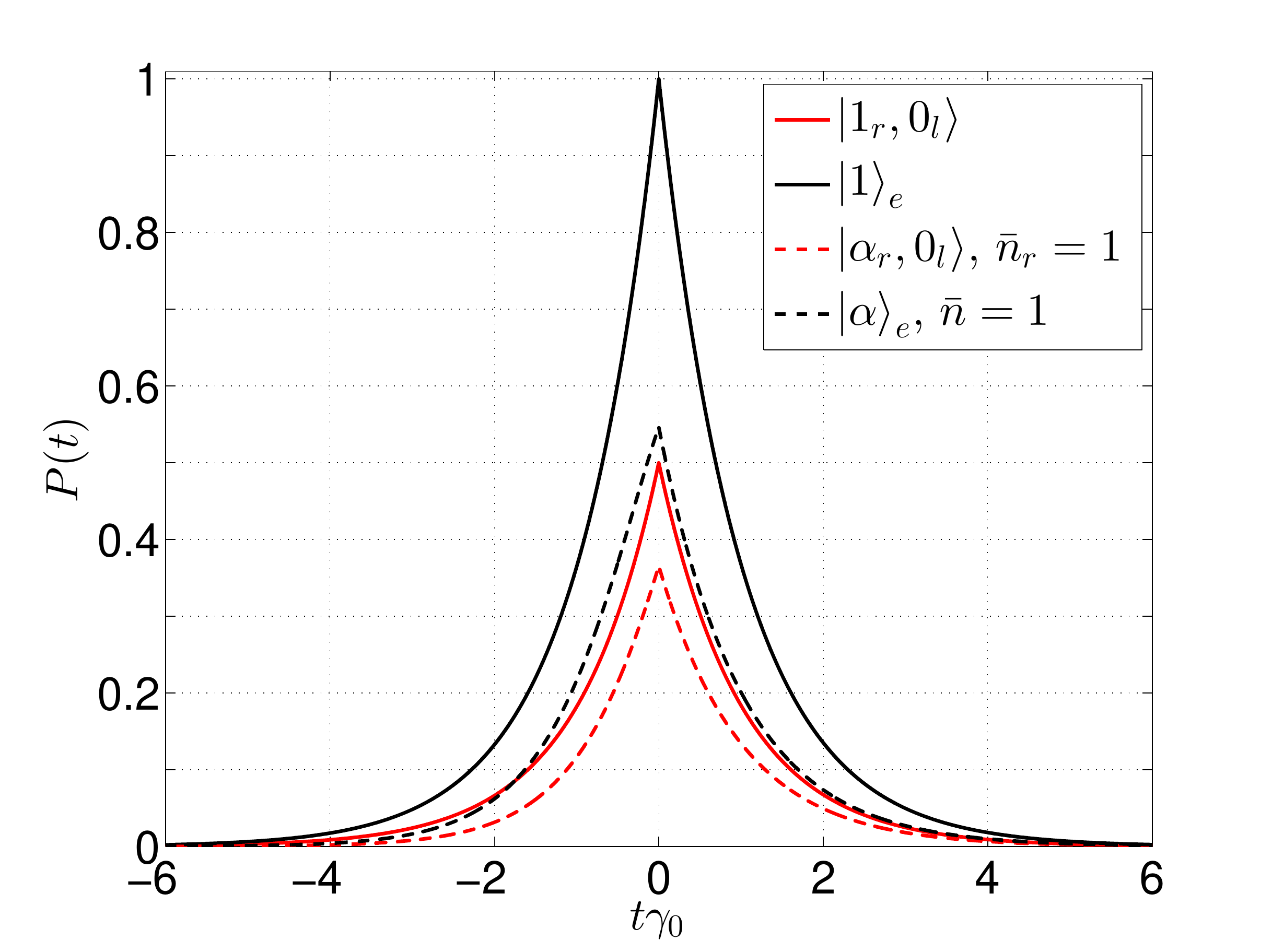}
\caption{(Color online) Atomic excitation probability $P(t)$ as a function of time with initial rising exponential pulses prepared in their optimum bandwidths for Fock state (solid curves) and coherent state (dashed curves) in the right-propagating mode (red curves) and the even-mode (black curves), respectively.}
\label{fig_pt_exp}
\end{figure}

For single-photon coherent state (dashed curves), the maximum excitation probability is $P_{max}=0.37$ for the photon being in a single spatial-mode $\ket{\alpha_r, 0_l}$ with the optimum bandwidth being $\Omega=1.36\gamma_0$ (red dashed curve) and $P_{max}=0.56$ for the photon being in the even-mode $\ket{\alpha_e}$ with $\Omega=1.9\gamma_0$ (black dashed curve). The atomic excitation $P(t)$ as a function of time for rising exponential pulse with optimum bandwidth is given in \fig{\ref{fig_pt_exp}}. One can see that for Fock and coherent state pulses, it takes almost the same time---on the order of atomic lifetime to achieve the maximum atomic excitation.


\subsection{Multi-photon excitation}
Here we consider case of $n$ photons in the even parity-mode interacting with the atom, schematically shown in \fig{\ref{fig_2_mode_model}}(c). We performed a numerical study for both Fock state and coherent state pulses with photon number ranging over $n\in[1,10]$.

\begin{figure}[h!]
\begin{minipage}{0.49\linewidth}
  \leftline{\includegraphics[width=4.34cm]{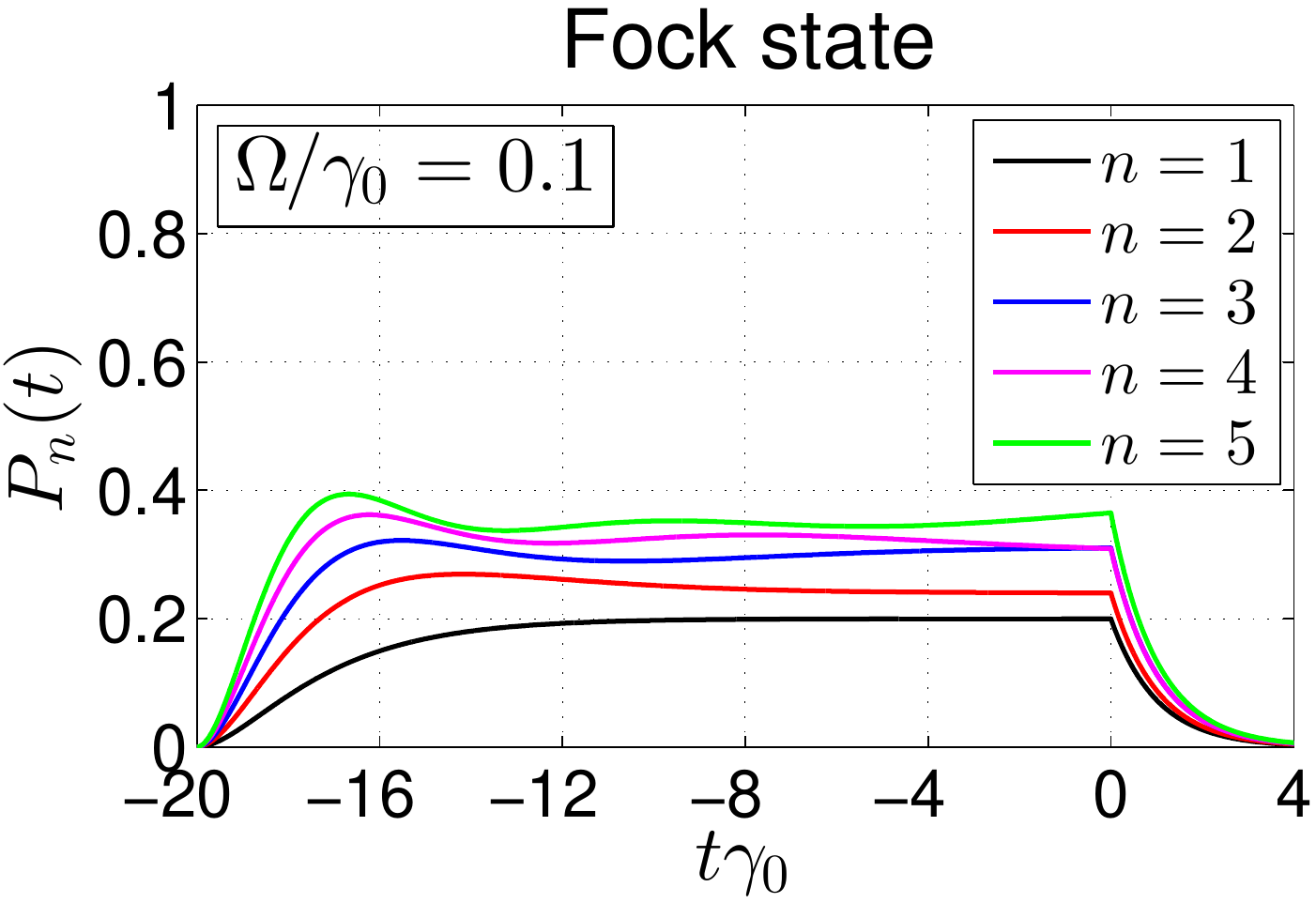}}
\vspace{0.0cm}
\end{minipage}
\hfill
\begin{minipage}{0.49\linewidth}
  \leftline{\includegraphics[width=4.34cm]{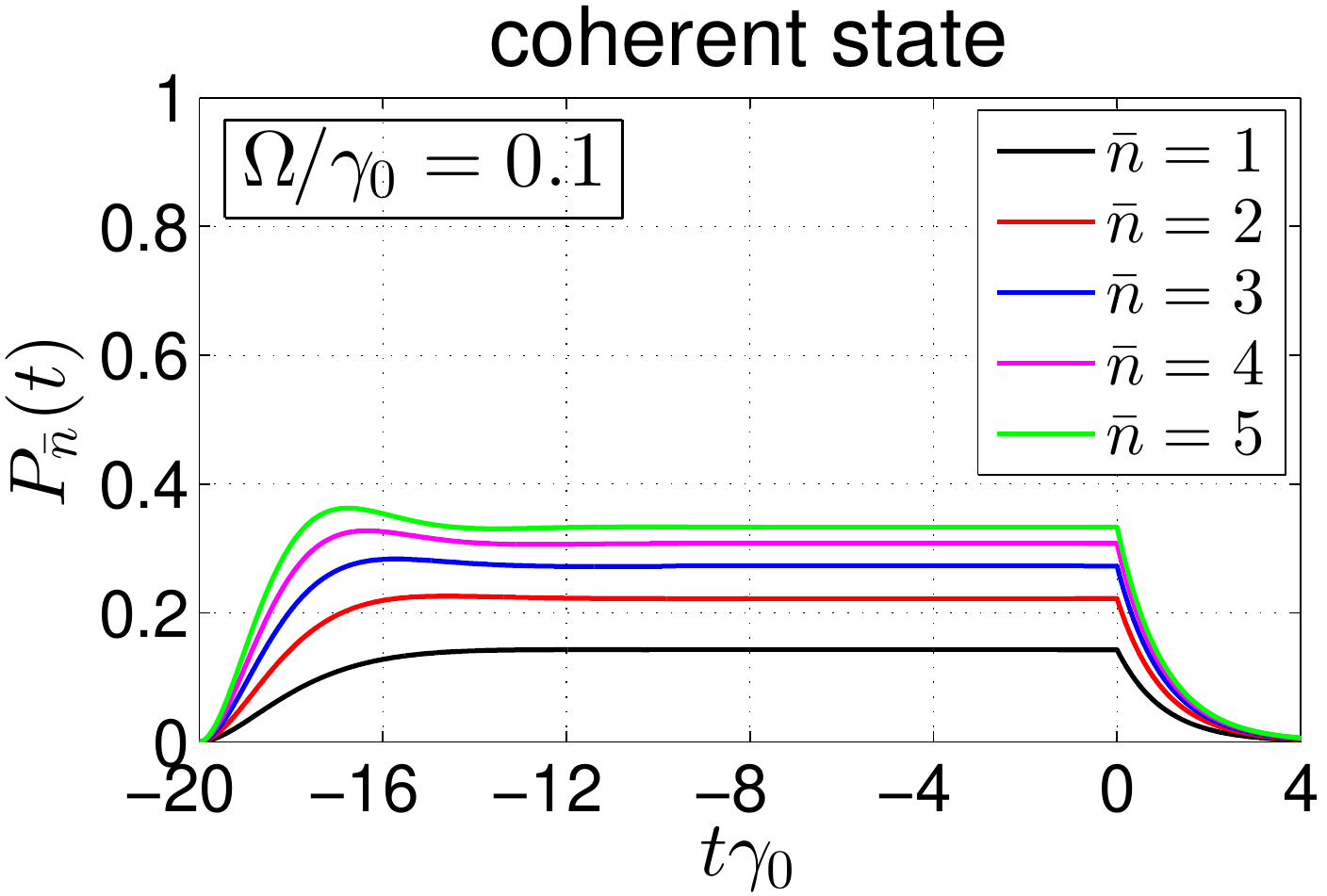}}
\vspace{0.0cm}
\end{minipage}
\begin{minipage}{0.49\linewidth}
  \leftline{\includegraphics[width=4.34cm]{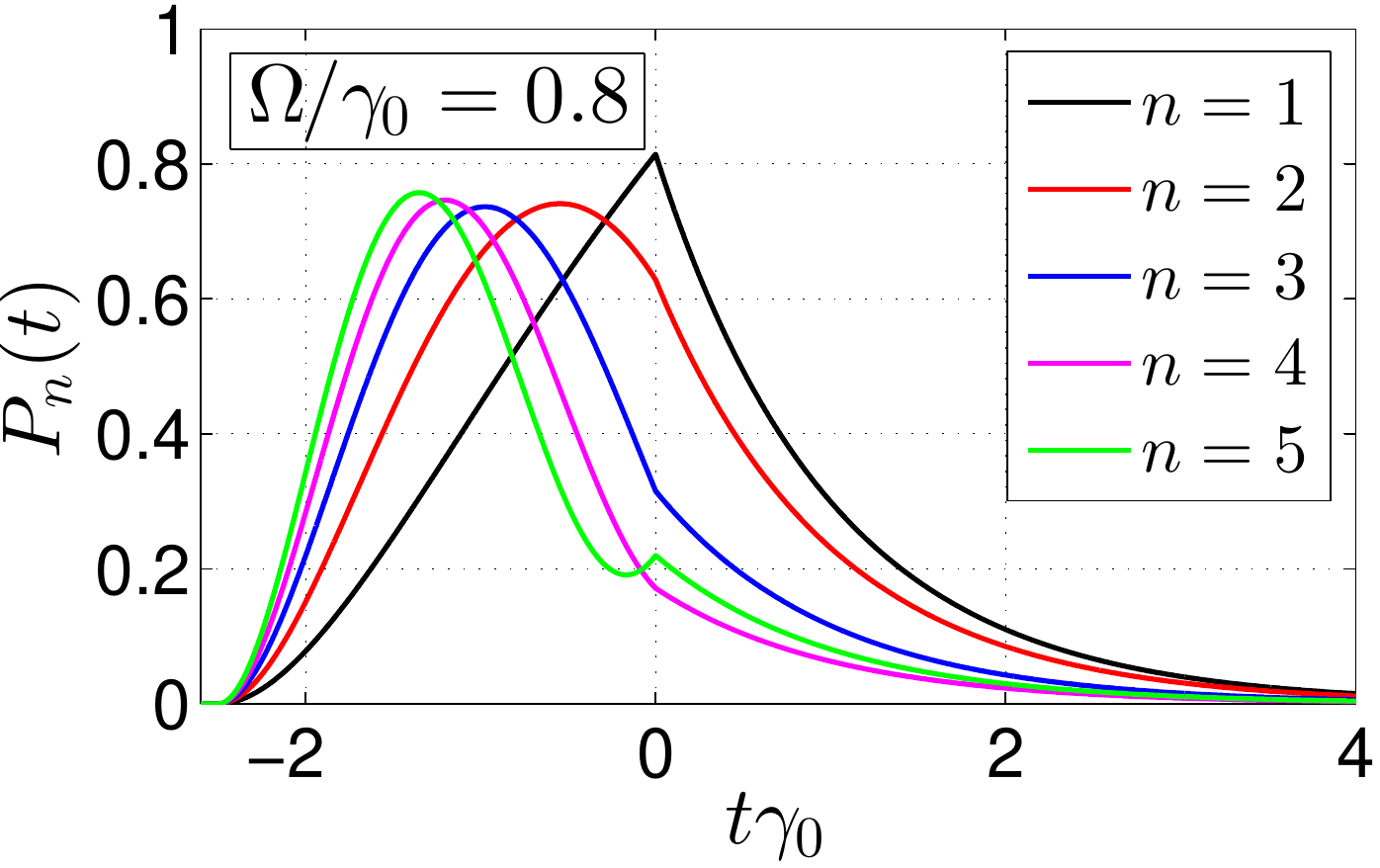}}
\vspace{0.0cm}
\end{minipage}
\hfill
\begin{minipage}{0.49\linewidth}
  \leftline{\includegraphics[width=4.34cm]{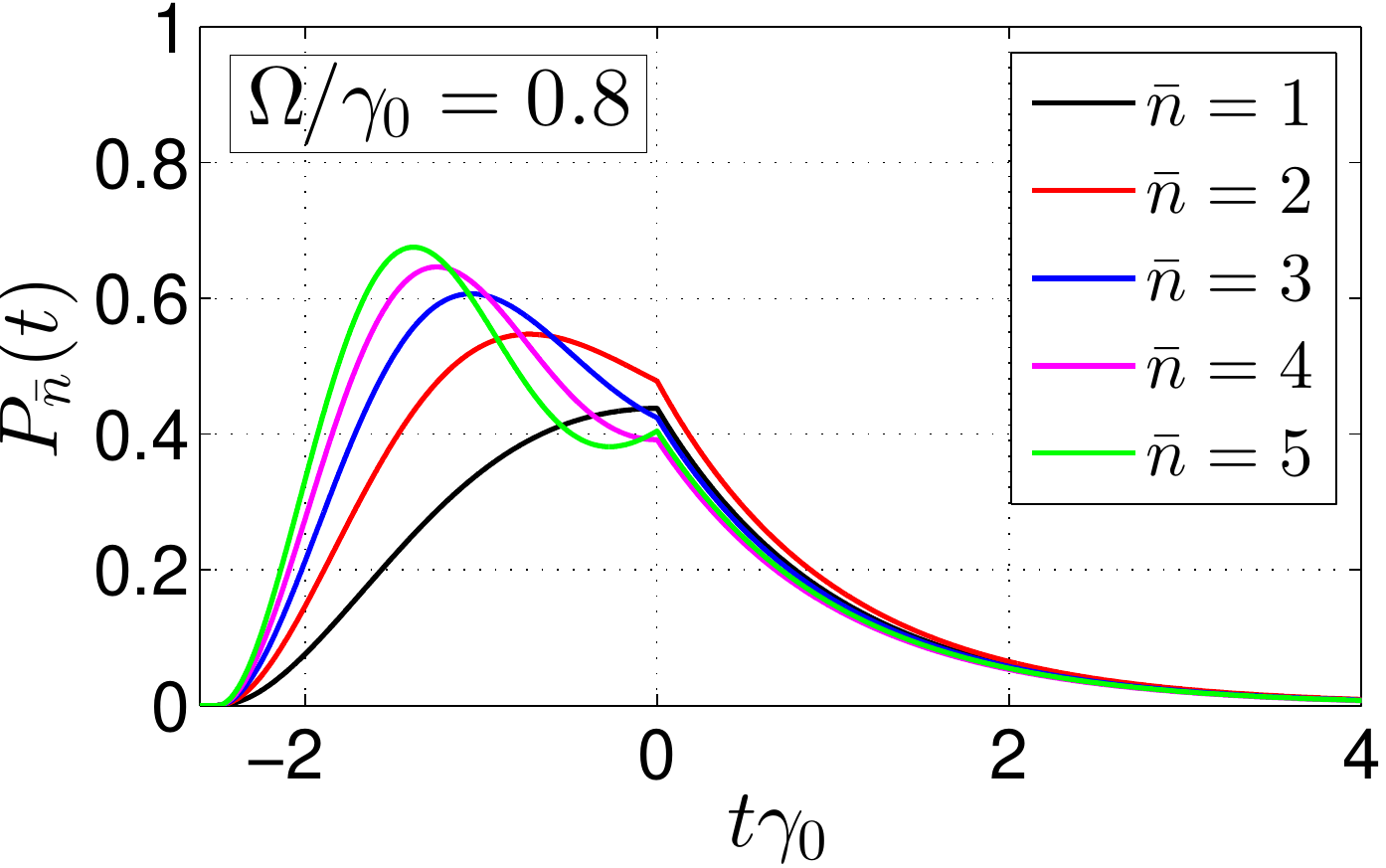}}
\vspace{0.0cm}
\end{minipage}
\begin{minipage}{0.49\linewidth}
  \leftline{\includegraphics[width=4.34cm]{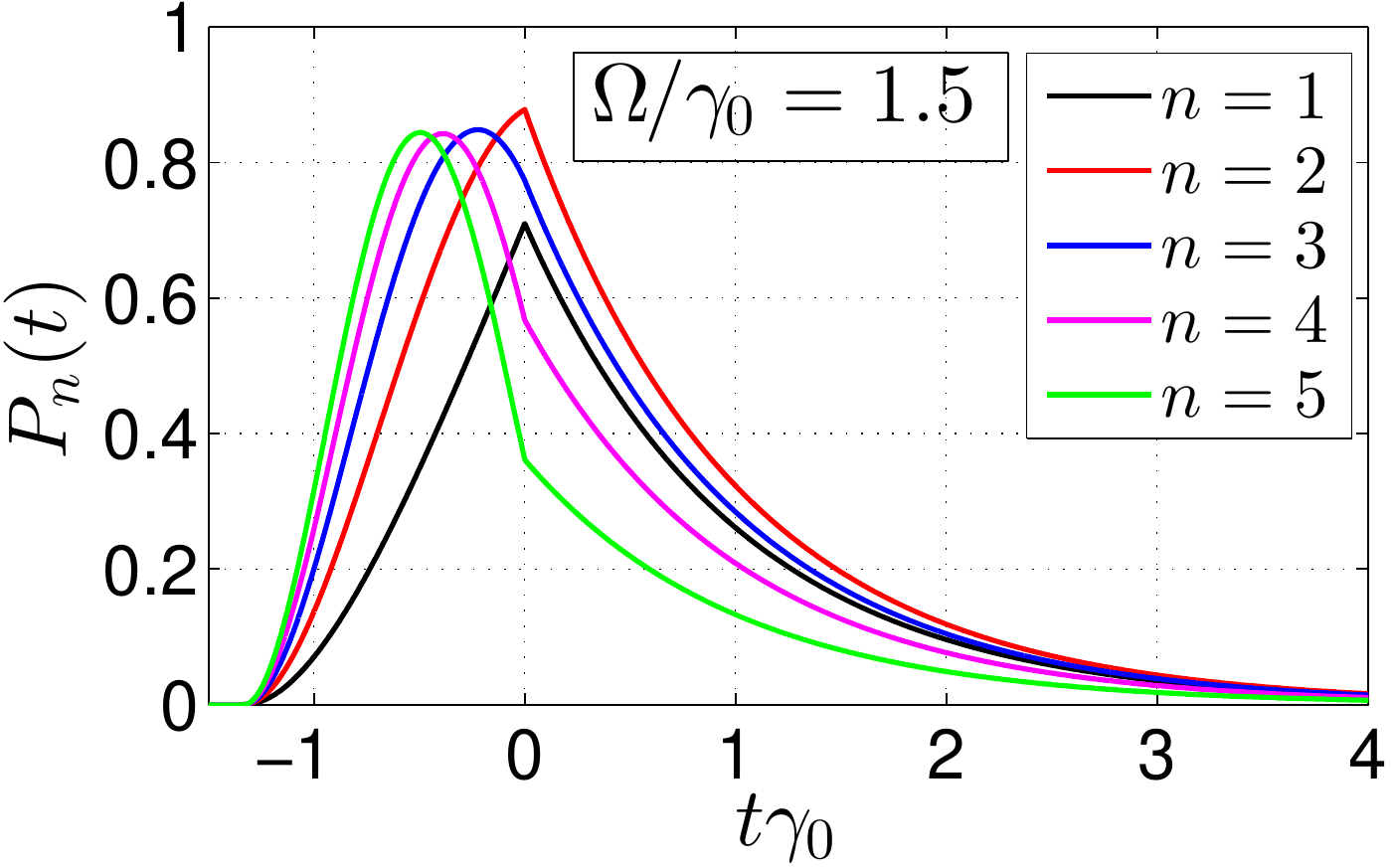}}
\vspace{0.0cm}
\end{minipage}
\hfill
\begin{minipage}{0.49\linewidth}
  \leftline{\includegraphics[width=4.34cm]{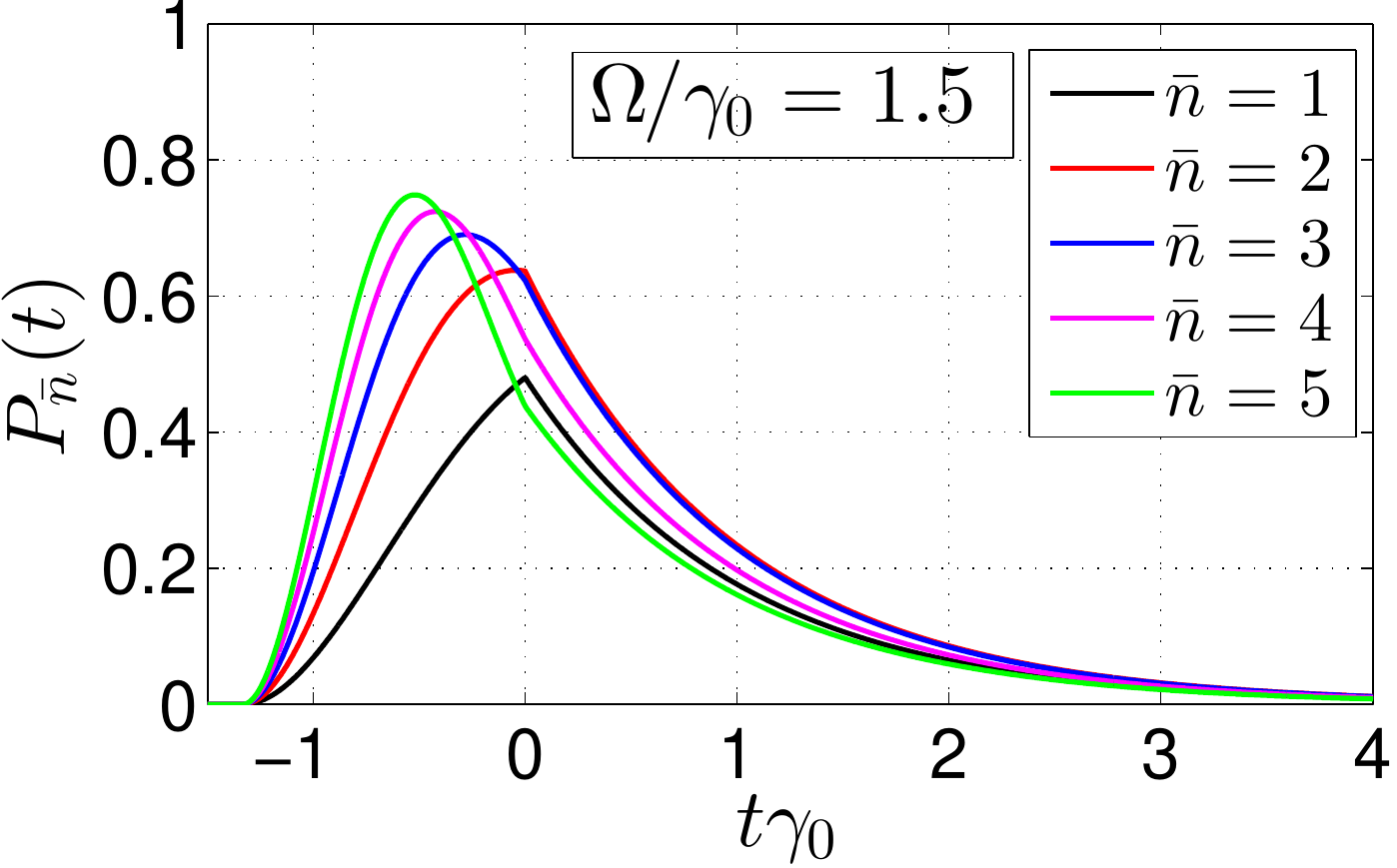}}
\vspace{0.0cm}
\end{minipage}
\begin{minipage}{0.49\linewidth}
  \leftline{\includegraphics[width=4.34cm]{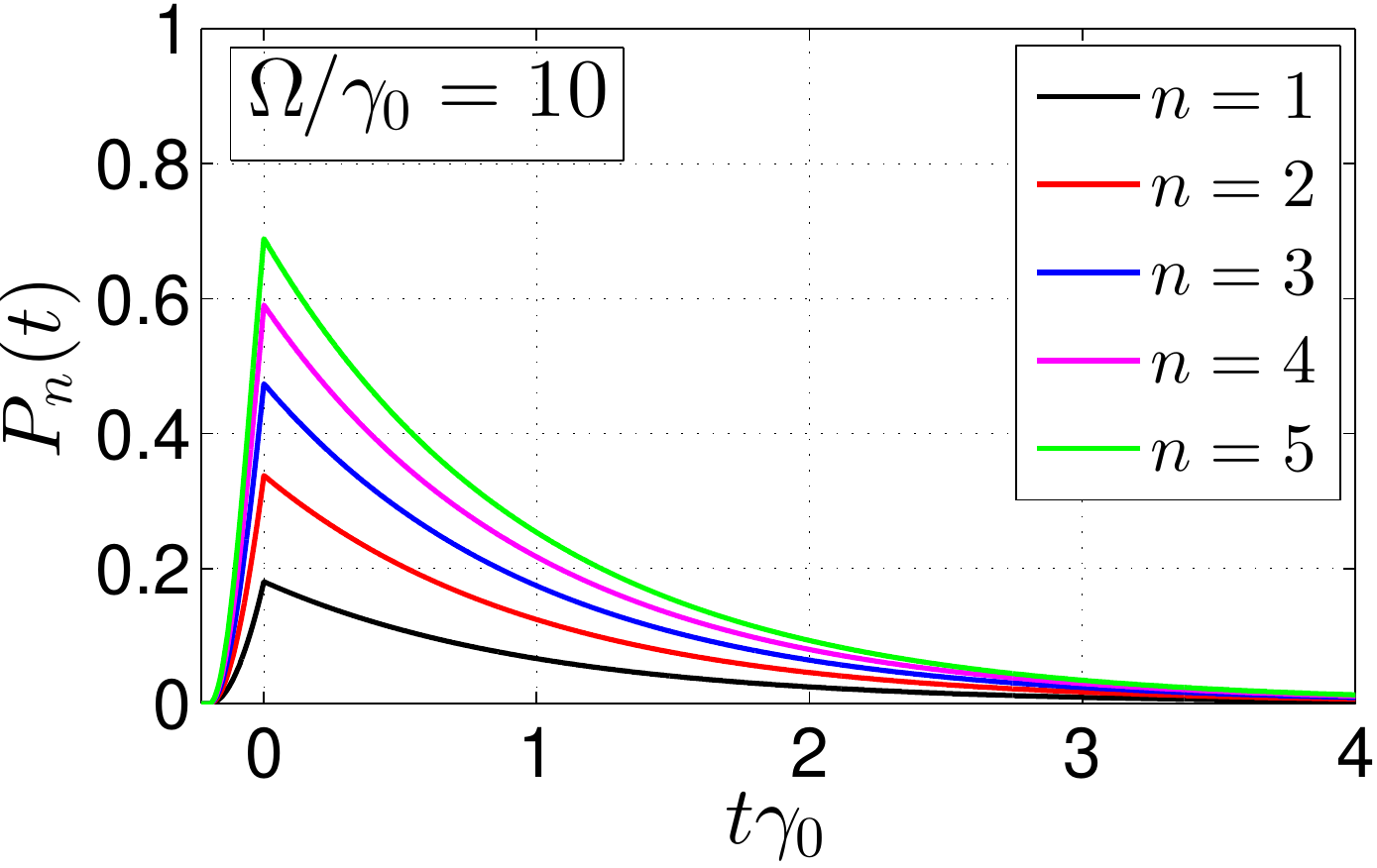}}
\vspace{0.0cm}
\end{minipage}
\hfill
\begin{minipage}{0.49\linewidth}
  \leftline{\includegraphics[width=4.34cm]{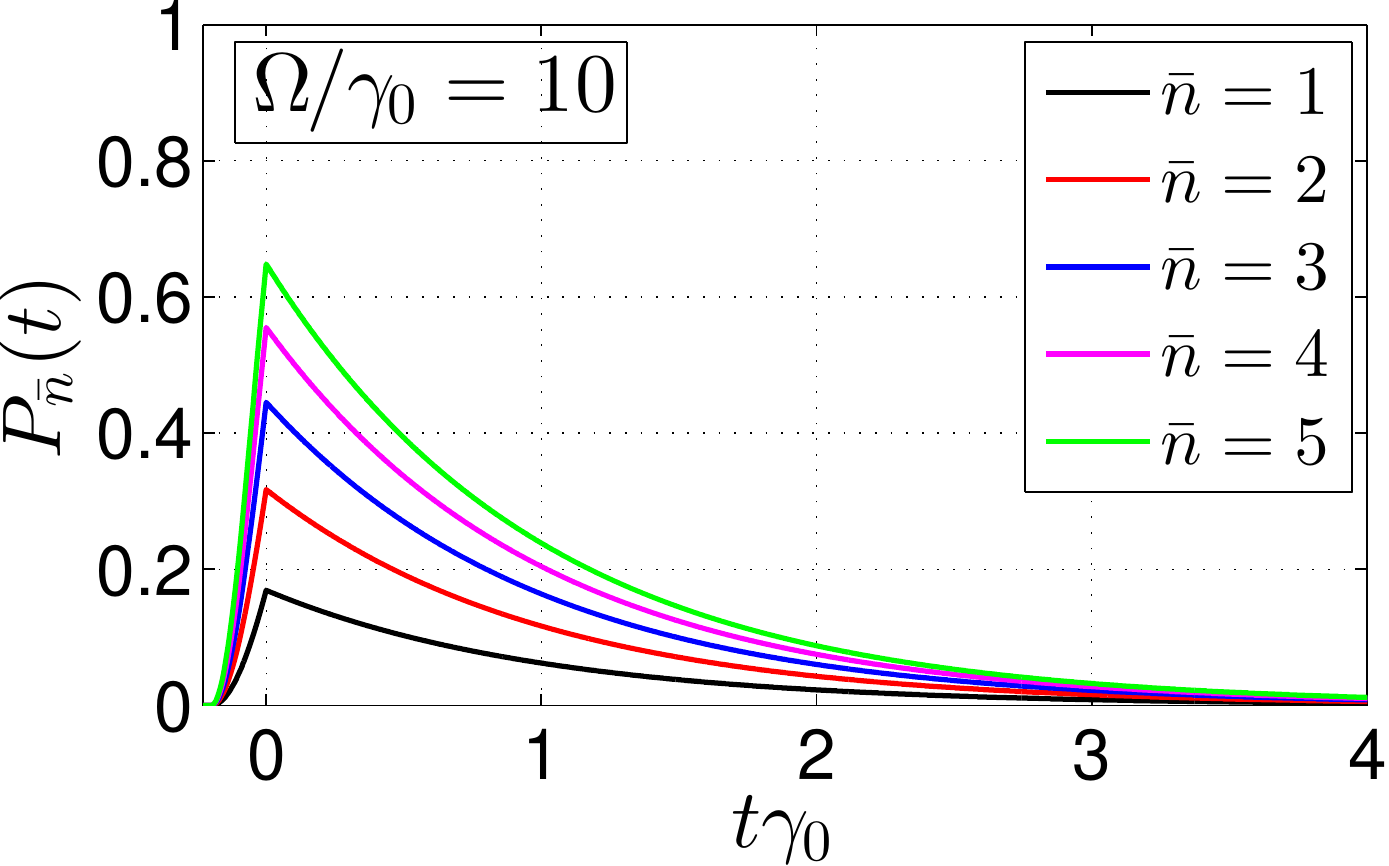}}
\vspace{0.0cm}
\end{minipage}
\caption{(Color online) Atomic excitation probability $P(t)$ as a function of time with the initial rectangular-shaped pulse for different bandwidths $\Omega/\gamma_0=\{0.1,0.8,1.5,10\}$ and different photon numbers $n, \bar{n}\in[1,5]$ in the even-mode. For coherent state pulses (right column), the maximum excitation probability $P_{max}$ is always ordered by average photon number $\bar{n}$ for arbitrary bandwidth. This is not the case for Fock state (left column) in general, namely for bandwidths $\Omega/\gamma_0=\{0.8,1.5\}$. The results for the Fock state are similar to Fig. 2 in Ref. \cite{Baragiola_2012}, where the shape of the pulse was taken to be Gaussian.}
\label{fig_pt_ne_square}
\end{figure}

\begin{figure}[h!]
\begin{minipage}{0.49\linewidth}
  \leftline{\includegraphics[width=4.34cm]{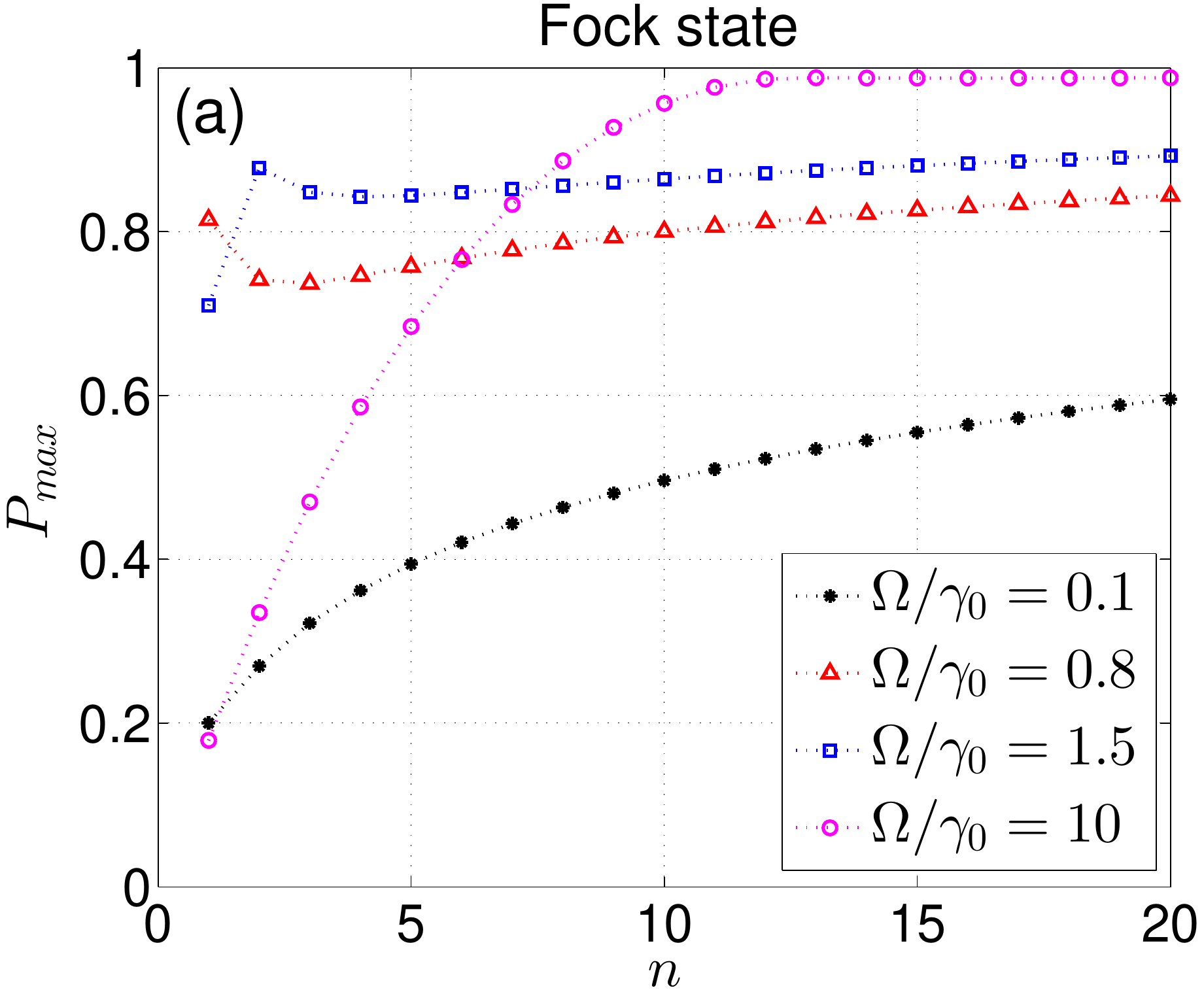}}
\vspace{0.0cm}
\end{minipage}
\hfill
\begin{minipage}{0.49\linewidth}
  \leftline{\includegraphics[width=4.34cm]{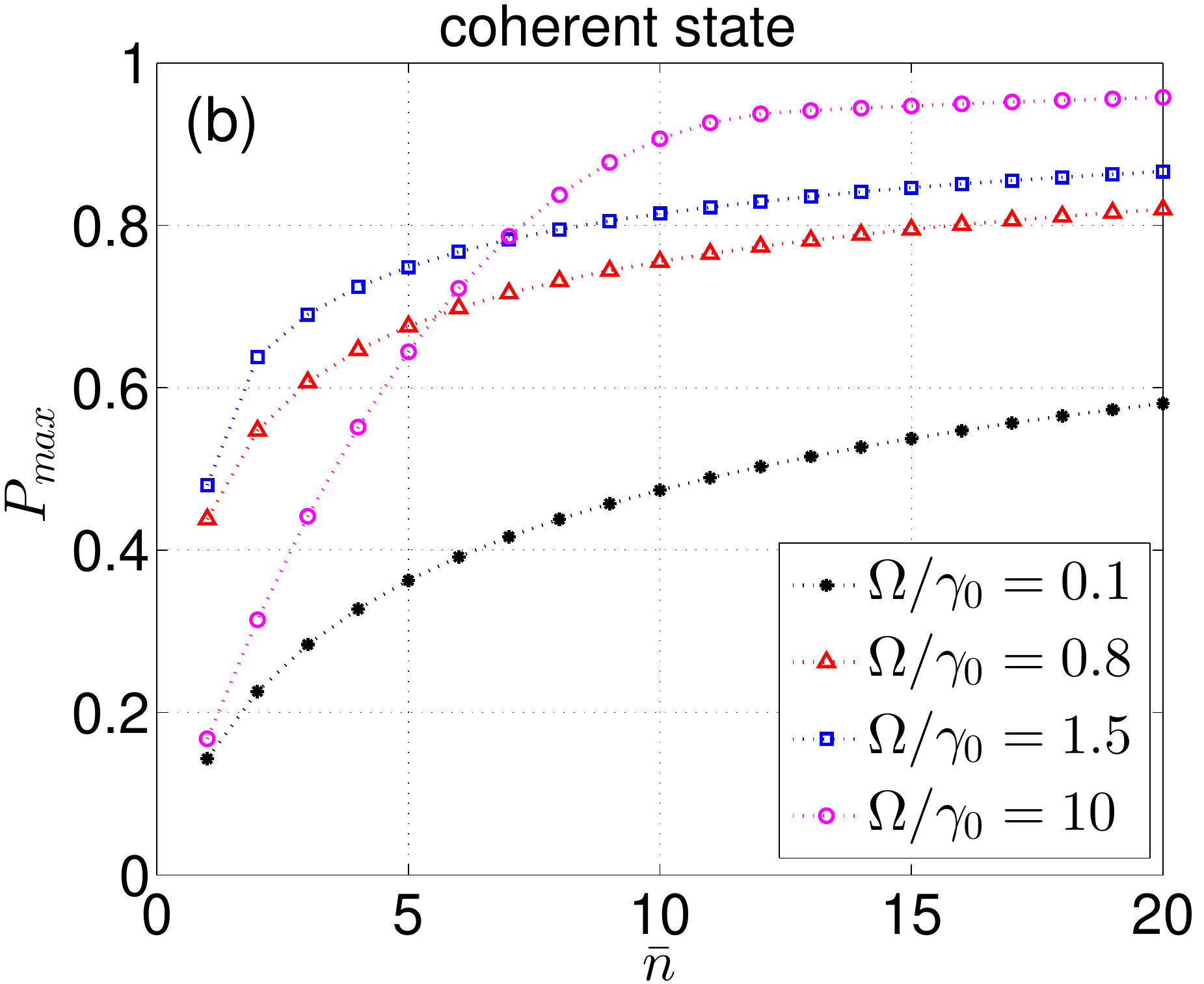}}
\vspace{0.0cm}
\end{minipage}
\caption{(Color online) Maximum excitation probability $P_{max}$ as a function of photon number with initial rectangular-shaped pulses in (a) Fock state (b) coherent state for different bandwidths $\Omega/\gamma_0=\{0.1,0.8,1.5,10\}$, respectively. $P_{max}$ increases monotonically with the photon number $\bar{n}$ for coherent state case (b). This is not the case for Fock state pulses with bandwidths $\Omega/\gamma_0=0.8\,(1.5)$, where there is a dip (a peak) at $n=2$.}
\label{fig_pmax_ne}
\end{figure}

In \fig{\ref{fig_pt_ne_square}}, excitation probability as a function of time with initial rectangular-shaped pulses in the even-mode for different bandwidths $\Omega/\gamma_0=\{0.1,0.8,1.5,10\}$ and with photon numbers $n=\{1,\cdot\cdot\cdot,5\}$ is plotted. As expected, better excitation is obtained with bandwidth close to the atomic linewidth $\Omega \approx \gamma_0$ and it takes shorter time to reach the maximum excitation for broader pulses. It is interesting to compare the Fock state cases (left column) and coherent state cases (right column) in \fig{\ref{fig_pt_ne_square}}. For bandwidth set $\Omega/\gamma_0=\{0.1,0.8,1.5,10\}$, maximum excitation probability $P_{max}$ is always ordered by photon number for coherent state pulses. But for Fock state pulses, $P_{max}$ is ordered by photon number only for narrow and broad bandwidths $\Omega/\gamma_0=\{0.1,10\}$, not for the intermediate bandwidths $\Omega/\gamma_0=\{0.8,1.5\}$.
For $\Omega/\gamma_0=0.8$, single-photon yields a better atomic excitation than higher photon number pulses and for $\Omega/\gamma_0=1.5$, two-photon excitation is better than the others. This effect can be seen more clearly from \fig{\ref{fig_pmax_ne}}, where the maximum excitation probability versus photon number for different bandwidths $\Omega/\gamma_0=\{0.1,0.8,1.5,10\}$ is plotted. For Fock state pulse with bandwidths $\Omega/\gamma_0=\{0.1,10\}$, $P_{max}$ increases monotonically with the photon number $n$, as shown in \fig{\ref{fig_pmax_ne}} (a). This is not true for $\Omega/\gamma_0=\{0.8,1.5\}$, where one can see a dip or peak at $n=2$ in the corresponding curves. For coherent state pulses shown in \fig{\ref{fig_pmax_ne}} (b), $P_{max}$ increases monotonically with the average photon number $\bar{n}$. It can also be seen from \fig{\ref{fig_pmax_ne}} that for both cases, atomic excitation increases faster for broader bandwidth ($\Omega/\gamma_0=10$) before saturation. This can be understood from the fact that in a short-pulse limit---pulse with duration far shorter than atomic lifetime, the spontaneous emission effect can be ignored. It is worth mentioning that Rabi oscillation can be seen with higher photon numbers in both cases.

\begin{figure}[h!]
\includegraphics[scale=0.3]{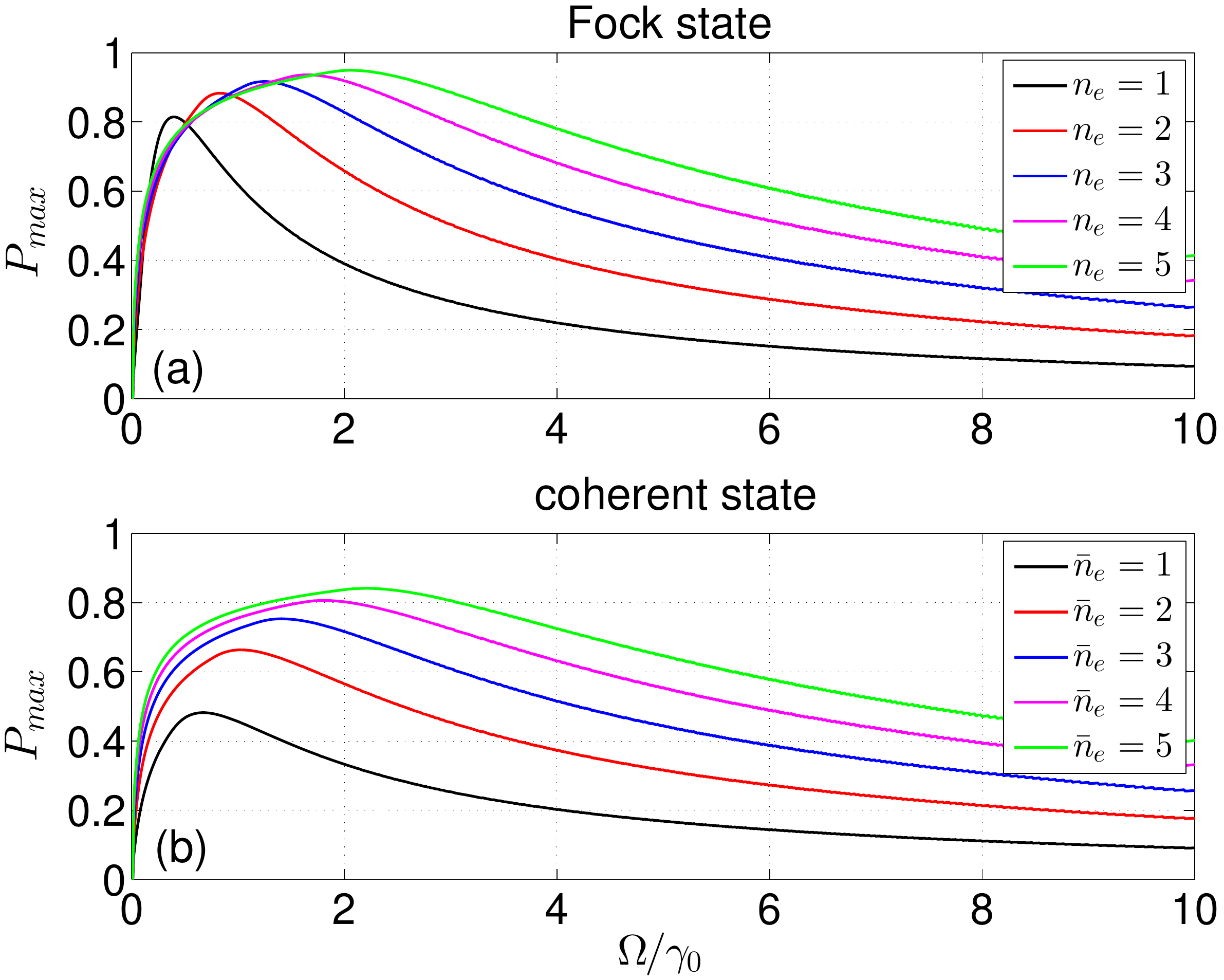}
\caption{(Color online) Maximum atomic excitation probability $P_{max}$ as a function of bandwidth $\Omega/\gamma_0$ with initially rectangular-shaped pulse in the even-mode for different photon numbers. The crossings in Fock state case (a) implies that for a given bandwidth there is a optimal photon number which maximizes the excitation probability. For coherent state pulses (b), higher photon number always yields higher $P_{max}$ for arbitrary bandwidths before saturation.}
\label{fig_pmax_ne_omega}
\end{figure}

We further study the optimum bandwidth for different photon-number pulses. Numerical simulation with rectangular-shaped pulse is shown in \fig{\ref{fig_pmax_ne_omega}} for $n\in[1,5]$. As expected, in the Fock state case \fig{\ref{fig_pmax_ne_omega}} (a), there are several crossings between lines for different photon numbers contrary to the coherent state case \fig{\ref{fig_pmax_ne_omega}} (b) with no line-crossings. For Fock state pulse with a given bandwidth, there is indeed a preferred photon number that maximizes the excitation probability and vice versa. But for coherent state pulses, higher photon number always gives higher maximal excitation probability before saturation.

\subsection{Interference of pulses in two distinct spatial-modes}
Let us consider two counter-propagating pulses incident simultaneously on a two-level atom, schematically shown in \fig{\ref{fig_2_mode_model}}(d). We consider the pulses being either in Fock state or in coherent state.

\subsubsection{Two-mode Fock state pulse}
\begin{figure}[h!]
\begin{minipage}{0.49\linewidth}
  \leftline{\includegraphics[width=4.34cm]{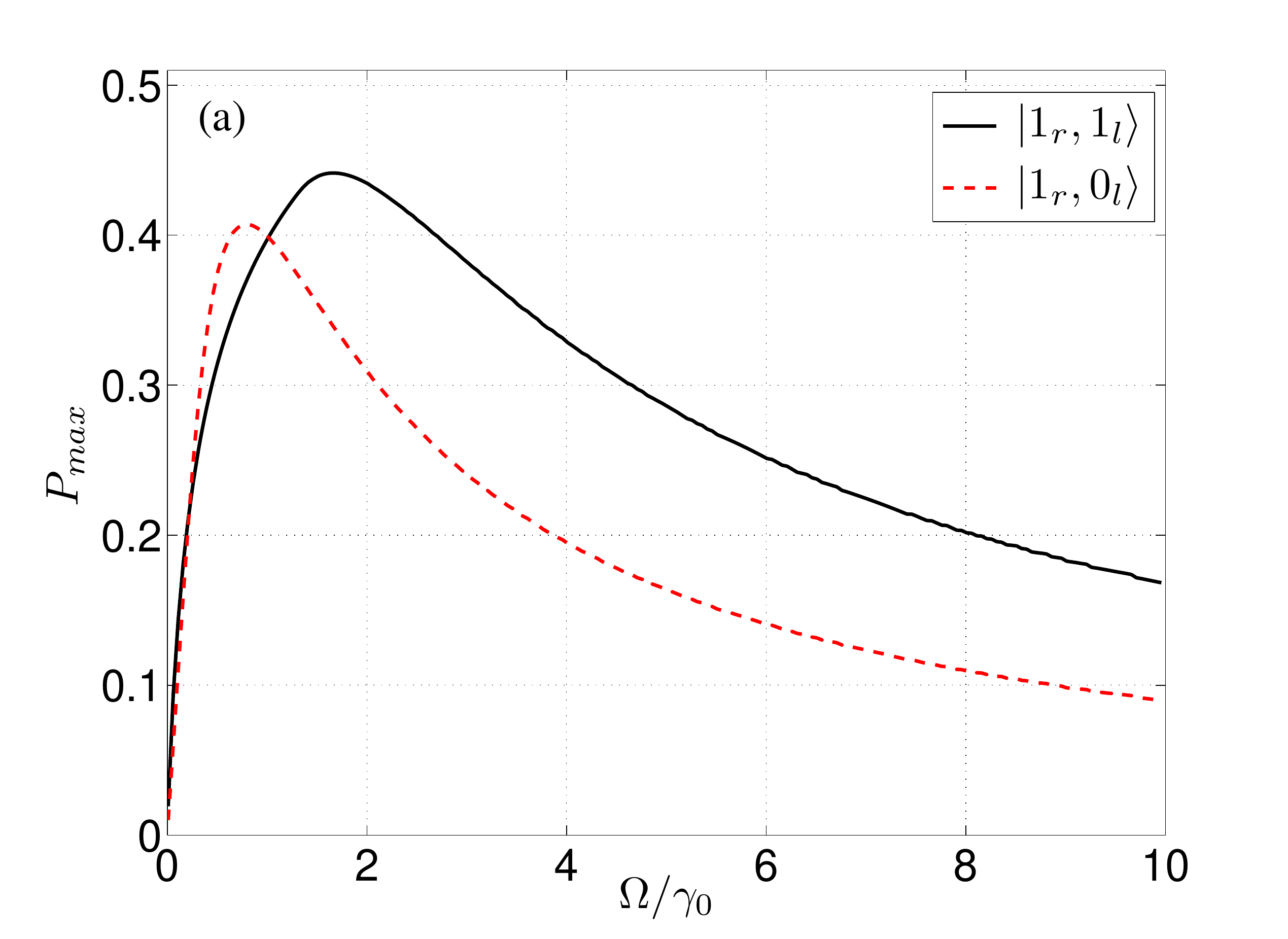}}
\vspace{0.0cm}
\end{minipage}
\hfill
\begin{minipage}{0.49\linewidth}
  \leftline{\includegraphics[width=4.34cm]{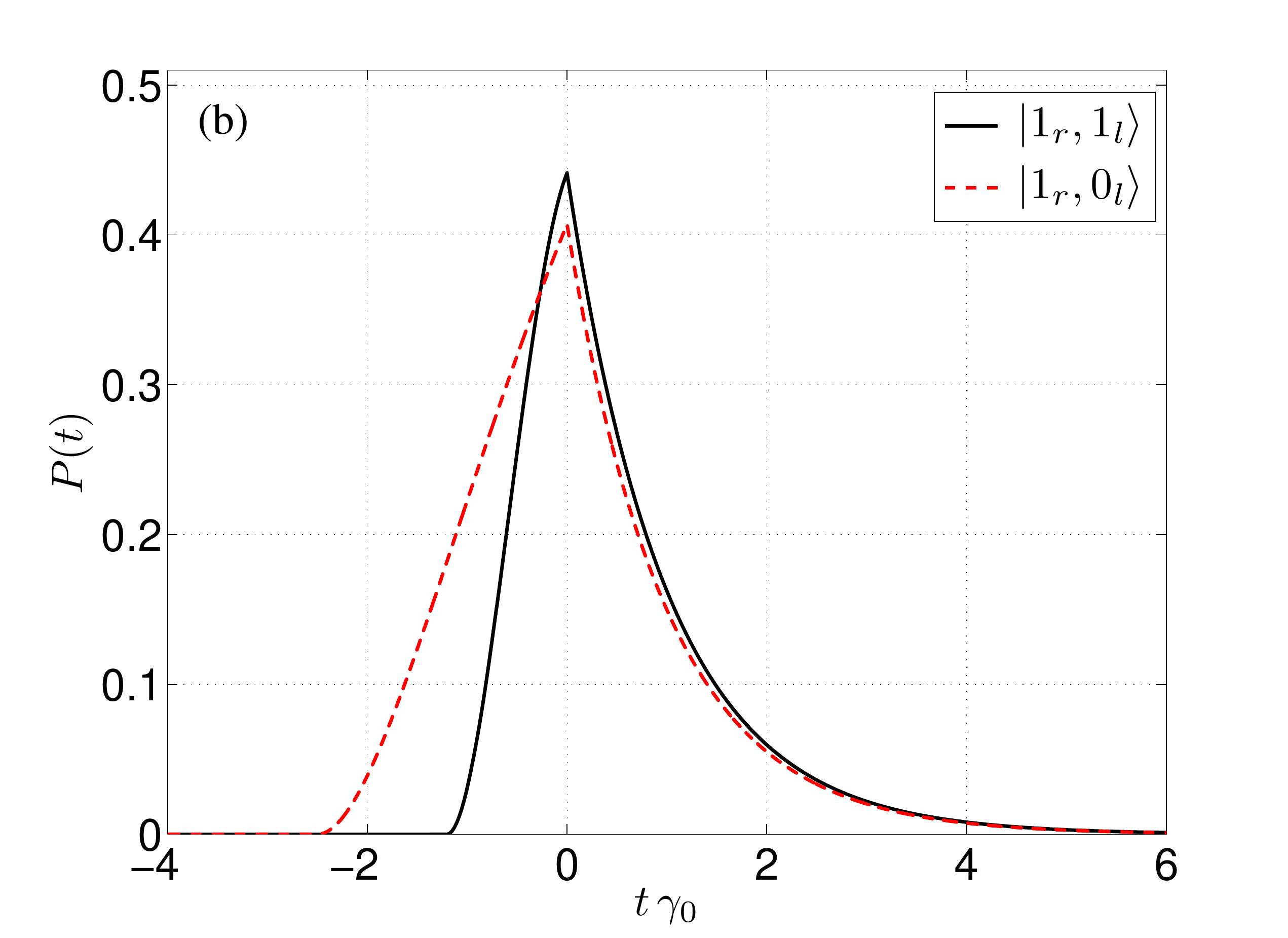}}
\vspace{0.0cm}
\end{minipage}
\caption{(Color online) Atomic excitation with initially rectangular-shaped Fock state pulses in different states. (a) dependence of $P_{max}$ on the pulse bandwidth. (b) time evolution of $P(t)$ for pulses with optimized bandwidth. The black solid line corresponds to the state $\ket{1_r, 1_l}$, and the red dashed line to the state $\ket{1_r, 0_l}$.}
\label{fig_f10_f11_square}
\end{figure}

Firstly, we consider the action of two single-photon Fock state pulses incident on the two-level atom. This corresponds to an initial state $\ket{\psi(t_0)}=\ket{g,1_r,1_l,0_s}$ of the total system. We assume the same rectangular temporal shape for the two pulses. The phase is completely undefined for a Fock state so that the constructive or destructive interference doesn't occur in this case \cite[p. 73]{Bachor_2004}. The dependence of atomic excitation on pulse bandwidth is given in \fig{\ref{fig_f10_f11_square}} (a), and the excitation probability  with time is given in \fig{\ref{fig_f10_f11_square}} (b). We can see that two counter-propagating pulses with identical rectangular shape give slightly better excitation than the single-photon with the same shape, but the excitation is still bounded by $0.5$---which we verified using two counter-propagating rising-exponentially shaped pulses with optimal bandwidth. This can be easily understood since
\beq
\ket{1_r,1_l} = \frac{\ket{2_e,0_o}-\ket{0_e,2_o}}{\sqrt{2}},
\eeq
which means there is a probability one half of no photon in the even-mode. This also explains the faster excitation by the state $\ket{1_r, 1_l}$ (\fig{\ref{fig_f10_f11_square}} (b)) due to the simultaneous presence of the two photons instead of one as in state $\ket{1_r, 0_l}$. It is worth mentioning that, the state
\beq
\frac{\ket{2_r,0_l}+\ket{0_r,2_l}}{\sqrt{2}}= \frac{\ket{2_e,0_o}+\ket{0_e,2_o}}{\sqrt{2}},
\eeq
gives the same atomic dynamics as the state $\ket{1_r,1_l}$, which is clear because those two states have the same even-mode component, which is the only component contributing to the atomic excitation.

\subsubsection{Two-mode coherent state pulse}
It is easy to recognize from \eqs{\ref{eq_dscn}} that for coherent state pulses, the appearance of the relative phase $\phi$ between the counter-propagating pulses gives rise to interference effects. To have better excitation of the atom, the counter-propagating pulses must interference constructively. In a particular case, for the two pulses with the same average photon number $\bar{n}_r=\bar{n}_l$ and relative phase difference $\phi=\pi$, the atom looks transparent to the two pulses, which propagate freely and wont't be affected by the atom.

\begin{figure}[h!]
\includegraphics[scale=0.3]{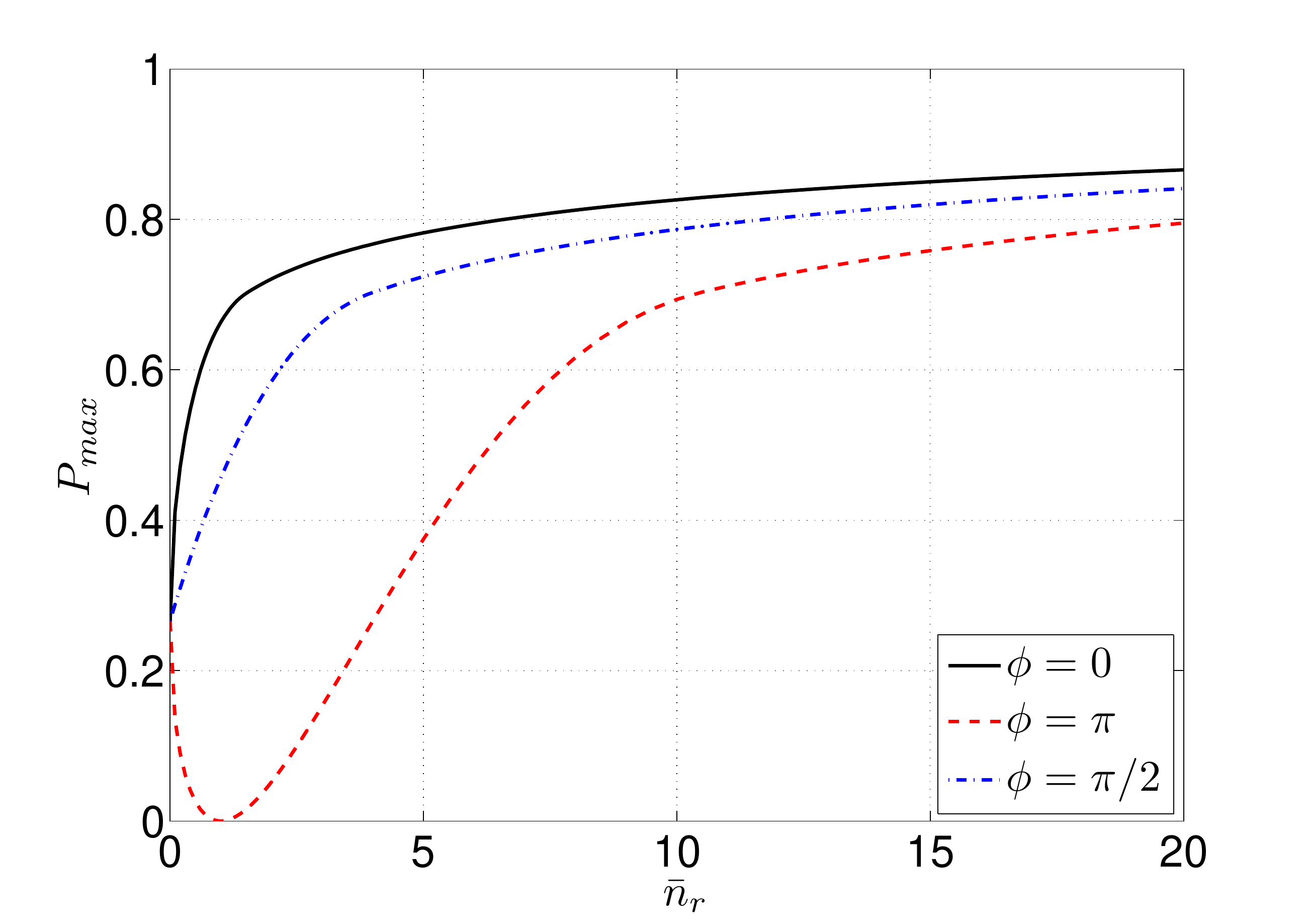}
\caption{(Color online) Maximum excitation probability $P_{max}$ as a function of mean photon number $\bar{n}_r$ in the right-propagating mode for different relative phases $\phi=\{0,\pi, \pi/2\}$ with average photon number one $\bar{n}_l=1$ in the left-propagating mode. The two pulses have the same rectangular shape with $\Omega=2\gamma_0$, which is the optimum bandwidth for $n_r=n_l=1$.}
\label{fig_pmax_nr}
\end{figure}
A simple application of the coherent state pulses interference effect is that the atomic excitation can be controlled by not only the mean photon number $\bar{n}_{r,l}$ but also the relative phase $\phi$. The maximum excitation probability $P_{max}$ varies as a function of the mean photon number $\bar{n}_r$ in the right-propagating mode for different relative phases $\phi=\{0,\pi, \pi/2\}$ as shown in \fig{\ref{fig_pmax_nr}} with $\bar{n}_l=1$. For two pulses with the same phase $\phi=0$ (black solid line), $P_{max}$ increases monotonically with the photon number in the right-propagating mode; for two pulses with the opposite phase $\phi=\pi$ (blue dashed line), a completely destructive interference happens for the same average photon number in the two spatial-modes $\bar{n}_r=\bar{n}_l=1$; for a phase difference of $\phi=\pi/2$ between the two pulses, a reduction of atomic excitation is observed (blue dash-dot line).

\section{Conclusion} 
\label{sec_con}
In conclusion, we have theoretically investigated the atomic dynamics due to the interaction with two spatial-mode multi-photon propagating pulses. Using a fully quantum mechanical treatment, the dependence of atomic excitation probability on different incident pulses have been studied and presented with general formalisms for both Fock state and coherent state. We have shown the following properties of atom dynamics in the one-dimensional two spatial-mode geometry:
\begin{itemize}
  \item {\it Single-photon excitation:} the atomic excitation probability is upper bounded by $0.5$, when the atom is excited by a single-photon from a single spatial-mode. Full atomic excitation by single-photon is possible only with a rising-exponentially shaped Fock state pulse in the even-parity mode---a balanced superposition of the right and left spatial-modes.
  \item {\it Multi-photon excitation:} for coherent state pulse, the maximum excitation probability $P_{max}$ is ordered by average photon number $\bar{n}$ in the even-mode for all bandwidths. Higher power always gives better atomic excitation before saturation. On the other hand, for Fock state pulses with intermediate bandwidths $\Omega\sim\gamma_0$, $P_{max}$ is not ordered by photon number $n$.
  \item {\it Two spatial-mode pulses interference:} in general, there is no interference between two Fock state pulses. On the other hand, for two coherent state pulses, the atomic dynamics can be well controlled by the relative phase $\phi$ between the two pulses and by the average photon number in both pulses.
\end{itemize}
These results are relevant for applications in integrated quantum optical devices, such as quantum switch for light \cite{Chang_2007, Mariantoni_2008}. In addition, the presented formalism can be used to further study the atom and propagating light pulses dynamics in one dimensional frequency-continuum. It can also be generalized to 3D cases when accounting for the details of the spatial mode-matching.

\section{Acknowledgements}
We would like to thank Colin Teo, Ben Q. Baragiola and Joshua Combes for useful discussions. This work was supported by the National Research Foundation and the Ministry of Education, Singapore.

\bibliographystyle{prsty}
\bibliography{qo_ref}
\end{document}